\documentclass[12pt]{iopart}

\usepackage{color}
\usepackage{epsf}
\usepackage{graphicx}
\usepackage{bm}
\usepackage{subfigure}

\begin{document}

\newcommand{\la}{\langle}
\newcommand{\ra}{\rangle}
\newcommand{\ua}{\uparrow}
\newcommand{\da}{\downarrow}
\newcommand{\rar}{\rightarrow}
\newcommand{\be}{\begin{equation}}
\newcommand{\ee}{\end{equation}}
\newcommand{\bea}{\begin{eqnarray}}
\newcommand{\eea}{\end{eqnarray}}
\global\long\def\av#1{\left\langle #1 \right\rangle }

\newcommand{\jay}{J}
\newcommand{\cnodot}{}
\newcommand{\nb}[1]{n^{b}}
\newcommand{\nf}[1]{n^{f}}
\newcommand{\cdop}[2]{c^{\dagger}_{#2} }           
\newcommand{\cop}[2]{c^{\phdagger}_{#2 }}
\newcommand{\bdop}[1]{b^{\dagger}_{} }           
\newcommand{\bop}[1]{b^{\phdagger}_{}}
\newcommand{\chdop}[1]{c^{\dagger}_{} }           
\newcommand{\chop}[1]{c^{\phdagger}_{}}
\newcommand{\Q}[1]{Q^{\phdagger}_{#1}}
\newcommand{\Qd}[1]{Q^{\dagger}_{#1}}
\newcommand{\phdagger}{\phantom{\dagger}}
\newcommand{\subs}[2]{ {#1}_{#2} }
\newcommand{\imag}{i}
\newcommand{\sups}[2]{ {#1}^{#2} }
\newcommand{\expectation}[1]{\langle \;  #1 \; \rangle}
\newcommand{\nh}{\hat{n}}
\newcommand{\rp}{{r'}}
\newcommand{\hopi}[1]{\hat{h}_{#1} }
\newcommand{\hop}{{h}}
\newcommand{\half}{{\textstyle \frac{1}{2}}}
\newcommand{\quarter}{{\textstyle \frac{1}{4}}}
\newcommand{\dmu}{{\delta_\mu}}
\newcommand{\down}{\downarrow}
\newcommand{\up}{\uparrow}
\newcommand{\Hhub}{ H}
\newcommand{\bS}{\bm{S}}
\newcommand{\bP}{\bm{P}}
\newcommand{\bQ}{\bm{Q}}
\newcommand{\teezero}{{T_0}}
\newcommand{\teeplus}{{T_1}}
\newcommand{\teeminus}{{T_{-1}}}
\renewcommand{\cdop}[2]{c^{\dagger}_{#1,\,#2}}           
\renewcommand{\cop}[2]{c^{\phdagger}_{#1,\, #2 }}
\renewcommand{\bdop}[1]{b^{\dagger}_{} }           
\renewcommand{\bop}[1]{b^{\phdagger}_{}}
\renewcommand{\chdop}[1]{{\hat{c}}^{\dagger}_{#1} }           
\renewcommand{\chop}[1]{{\hat{c}}^{\phdagger}_{#1}}
\renewcommand{\nb}[1]{n^{b}_{}}
\renewcommand{\nf}[1]{n^{f}_{}}
\newcommand{\gee}[2]{ q^{#1}_{#2 } }
\newcommand{\bgee}[1]{ {\bm{q} }_{\, #1}   }


\title{
Spinon and $\eta$-spinon correlation functions
}

\author{P. D. Sacramento$^{1,2}$, Y. C. Li$^2$, S. J. Gu$^3$ and J.M.P. Carmelo$^{2,4,5}$ 
}
\address{$^1$ Centro de F\'{\i}sica das Interac\c{c}\~oes Fundamentais, Instituto Superior
T\'ecnico, Universidade T\'ecnica de Lisboa, Av. Rovisco Pais, 1049-001 Lisboa, Portugal\\
$^2$ Beijing Computational Science Research Center, Beijing 100084, China\\
$^3$ Department of Physics and Institute of Theoretical Physics,
The Chinese University of Hong Kong, China \\
$^4$ GCEP-Centre of Physics, University of Minho, Campus Gualtar, P-4710-057 Braga, Portugal \\
$^5$ Institut f\"ur Theoretische Physik III, Universit\"at Stuttgart, D-70550 Stuttgart, Germany}

\date{\today}

\begin{abstract}
We calculate real-space static correlation functions related to basic entities
of the one-dimensional Hubbard model, which emerge from the exact Bethe-ansatz
solution. These entities involve complex rearrangements of the original electrons.
Basic ingredients are operators related to unoccupied, singly occupied with spin up or spin down and
doubly occupied sites. The spatial decay of their correlation functions is determined
using an approximate mean-field-like approach based on the Zou-Anderson transformation
and DMRG results for the half-filled case. The nature and spatial extent of the correlations
between two sites on the Hubbard chain is studied using the eigenstates and eigenvalues of the
two-site reduced density matrix.  
\end{abstract}
\vspace{0.3cm}

\section{Introduction}

It was recently shown in \cite{npbIII} that a consistent description of the
Bethe-ansatz exact eigenstates of the Hubbard chain can be achieved
in terms of some basic entities called $c$ fermions, spinons and $\eta$-spinons.
(In the related preliminary studies of \cite{Carmelo_2004} such objects were
named $c$ pseudoparticles, spinons and holons, respectively.)
These objects are related to the rotated-electron occupancy configurations. The 
rotated-electron creation and annihilation operators are related to those of the
original electrons by a unitary transformation, $V(U)$. This transformation is 
defined such that double occupancy of these
rotated electrons is a good quantum number for all values of $U$.
There are infinite choices for such transformations. Examples are those
reported in \cite{Harris_1967,Stein_1997}. However, the BA solution
performs a specific electron - rotated-electron unitary transformation \cite{npbIII}.

The operator formulation introduced in \cite{npbIII} accounts for all representations
of the model global $[SO(4)\otimes U(1)]/Z_2$ symmetry algebra. Here 
$SO (4) =[SU(2)\otimes SU(2)]/Z_2$ refers to the spin $SU(2)$ and
$\eta$-spin $SU(2)$ symmetries and $U(1)$ to the $c$ hidden $U(1)$ symmetry
found in \cite{bipartite}. We denote the spin and the
$\eta$-spin of an energy eigenstate by $S_s$ and $S_{\eta}$, respectively.
We call $2S_c$ the number of rotated-electron singly occupied sites,
which is the eigenvalue of the generator of the $c$ hidden $U(1)$ symmetry.

At very large $U$ double occupancy is a good quantum number, since
there is a very large energy separation between states differing
by the number of doubly occupied sites (in this limit the unitary
transformation is the identity). As the Coulomb repulsion
becomes finite, the unitary transformation is not known explicitly
but it can be expressed in a perturbative expansion in powers of
$t/U$, as shown for instance in \cite{macdonald}. To leading order in $t/U$ 
the unitary operators associated with such transformations have 
a universal form. Such operators differ in their higher-order terms. 
Recently, the matrix elements of the unitary operator associated with
the specific transformation performed by the Bethe-ansatz solution 
has been obtained for the whole $U/t>0$ range in the basis of the energy 
eigenstates \cite{npbIII}.

The various entities referred to above are introduced so that the number of $c$ fermions
$N_c=2S_c$ and that of $c$ fermion holes $N_c^h =[N_a -2S_c]$
are equal to the number of singly occupied sites of rotated electrons (with spin up or spin
down) and the number of rotated-electron doubly and unnocupied sites, respectively.
They store information on the charge part of these rotated electrons.
The number of spinons $M_s = 2S_c$ also equals that of singly occupied sites of rotated electrons.
The spin-$1/2$ spinons of component $+1/2$ have information on the spin
of these rotated electrons of spin up, the spin-$1/2$ spinons of
component $-1/2$ is related to the spin part of such rotated electrons with spin down. 
The number of $\eta$-spinons $M_{\eta} = [N_a -2S_c]$ equals that of 
rotated-electron doubly and unnocupied sites. While the $N_c^h =[N_a -2S_c]$
$c$ fermion holes describe the $c$ hidden $U(1)$ symmetry degrees of
freedom of such sites occupancies, the $M_{\eta} = [N_a -2S_c]$ $\eta$-spinons 
refer to the $\eta$-spin $SU(2)$ symmetry degrees of freedom of the same
site occupancies. Specifically, the $\eta$-spin-$1/2$ $\eta$-spinons of component $+1/2$ are
related to the unoccupied occupied sites and the $\eta$-spin-$1/2$ $\eta$-spinons of component
$-1/2$ are related to the doubly sites of the rotated electrons.

Furthermore in \cite{npbIII,Carmelo_2004} it was proposed that both the
energy eigenstates inside and outside the Bethe-ansatz solution subspace
can be generated by occupancy configurations of the $c$ fermions (associated with
the usual real charge rapidities of the Bethe ansatz solution) and suitable spinon
and $\eta$-spinon occupancy configurations. 

Out of the $M_s = 2S_c$ spinons, a number $M_s^{un}=2S_s$ of spinons are unbound and determine the
energy eigenstate spin value $S_s$. For the energy eigenstates inside the Bethe-ansatz 
subspace, all unbound spinons have spin projection $+1/2$. Flipping such unbound spinons 
generates the spin towers of spin projection $S_s^z=-S_s,-S_s+1,...,S_s-1,S_s$. 
The corresponding energy eigenstates are outside the Bethe-ansatz subspace.
The remaining $M_s^{bo}=[2S_c- 2S_s]$ spinons are bound within composite entities with total spin zero
(bound-state between a spin-$1/2$ spinon with component $1/2$ and another spin-$1/2$ spinon with component $-1/2$).
There are also Bethe-anstaz excited states containing $\nu =1,2,3,...$ spin-neutral pairs
of such bound spinons. Note that zero-magnetization ground states have no unbound spinons, so
that all their spinons are bound within spin-neutral pairs. In this case one has that $M_s = 2N_{s1}$,
where $N_{s1}$ is the number of spin-singlet two-spinon composite objects called in
\cite{npbIII} $s1$ fermions.

Similarly, out of the $M_{\eta} = [N_a -2S_c]$ $\eta$-spinons, a number $M_{\eta}^{un} = 2S_{\eta}$ of $\eta$-spinons 
are unbound and determine the energy eigenstate $\eta$-spin value $S_{\eta}$. For the energy eigenstates 
inside the Bethe-ansatz subspace, all unbound $\eta$-spinons have $\eta$-spin projection $+1/2$. 
Those correspond to rotated-electron unoccupied sites. Flipping such unbound $\eta$-spinons 
generates the $\eta$-spin towers of $\eta$-spin projection $S_{\eta}^z=-S_{\eta},-S_{\eta}+1,...,S_{\eta}-1,S_{\eta}$.
Such $\eta$-spin flipping processes involve creation of on-site spin-neutral electron pairs of momentum $\pi$.
The corresponding energy eigenstates are outside the Bethe-ansatz subspace.
The remaining $M_{\eta}^{bo} = [N_a -2S_c-2S_{\eta}]$ $\eta$-spinons are anti-bound within composite entities 
with total $\eta$-spin zero (anti-bound state between a $\eta$-spin-$1/2$ $\eta$-spinon with component
$1/2$ and another one with component $-1/2$). Again, there are as well Bethe-anstaz excited states 
containing $\nu =1,2,3,...$ $\eta$-spin-neutral pairs of such anti-bound $\eta$-spinons. Each pair
involves two sites, doubly occupied and unoccupied by rotated electrons, respectively.

The eigenvalue $2S_c$ of the operator that counts the number of rotated-electron
singly occupied sites obeys the inequality $2S_c\leq N$. As a simple example
let us consider ground states with electronic density $n=N/N_a$ and spin
density $m=[N_{\uparrow}-N_{\downarrow}]/N_a$ in the ranges $n\in (0,1)$ and $m\in (0,n)$,
respectively. For such ground states one has that $2S_c=N$.  Within the operator formulation of \cite{npbIII}, those have 
$M_s = [M_{\eta}^{un} + M_{\eta}^{bo} ] = N$ spinons of which $M_{s}^{un} = 2S_{s}$
are unbound spinons and $M_{s}^{bo} = 2N_{s1}$ are bound spinons inside $N_{s1}= [N/2-S_s]$
spin-neutral two-spinon composite $s1$ fermions. Furthermore, such ground states have
$M_{\eta} = 2S_{\eta} = [N_a - N]$ $\eta$-spinons, $N_c = N$ $c$ fermions, and $N_c^h = [N_a - N]$
$c$ fermion holes. For them the number of electrons $N$ equals
that of rotated electrons that singly occupy sites and the number of rotated-electron doubly occupied
sites vanishes. The ground-state $M_s =N$ spinons refer to the $N$ spin-$1/2$ spins of the rotated electrons
that singly occupy sites. The ground-state $N_c = N$ $c$ fermions describe the charge degrees
of freedom of such rotated electrons. The ground-state $M_{\eta} = [N_a - N]$ $\eta$-spinons describe
the $\eta$-spin degrees of freedom of the $ [N_a - N]$ sites unoccupied by rotated electrons.
The ground-state $N_c^h = [N_a - N]$ $c$ fermion holes describe
the $c$ hidden $U(1)$ symmetry degrees of freedom of the latter $[N_a - N]$ sites ground-state occupancies.

Consistent with the results briefly reported above,
within the formulation of \cite{npbIII} the $\eta$-spin degrees of freedom of the rotated-electron unoccupied sites 
with component $1/2$ and the rotated-electron doubly occupied sites with component $-1/2$ tend to be 
anti-bound, whereas the singly occupied sites by electrons of opposite spin projection tend to be bound. 
One expects therefore correlations that should decrease somewhat fast with distance between the
members of each pair of $\eta$-spinons or of spinons.

The relevance of the correlations between doubly occupied sites and unoccupied
sites has been suggested before by several authors. This can involve the introduction of 
an effective low-energy theory that  contains a charge 2e bosonic mode \cite{Leigh,Choy,Phillips}, 
which may be bound to a hole. The significance of short-range correlations between unoccupied 
and doubly occupied sites was, for instance, found in \cite{Kaplan}.
The proposal of (anti-)bound states of doubly occupied sites and unoccupied sites
in the repulsive Hubbard model at half-filling (Mott insulating phase) and of
spins of opposite projections in the negative case (Luther-Emery phase), were
recently justified by the existence of long range-order in a non-local
order parameter \cite{Montorsi}.

The unitary transformation, $V(U)$, is such that $[\tilde{D},H]=0$, where
$\tilde{D}=V D V^{-1} = V D V^{\dagger}$, with $D$ the operator that
counts the number of doubly occupied sites of the original electrons.
At very large $U$, the eigenstates of the Hamiltonian may be labelled
by the eigenvalue of $D$ and at finite $U$ they may be labbelled by that of $\tilde{D}$. The rotated
electrons (and in general any operator written in terms of the rotated electrons)
can be obtained in the form $\tilde{c}=V c V^{-1}$, $\tilde{c}^{\dagger} = V c^{\dagger} V^{-1}$.
The eigenstates of the Hamiltonian at any value of $U$ may be generated as,
\be
|\psi_U \rangle = V |\psi_{U=\infty} \rangle
\ee
which uniquely defines the unitary transformation \cite{npbIII}. 
Therefore any correlation function of two operators $A$ and $B$ satisfies,
\bea
\langle \psi_{U=\infty} | A B | \psi_{U=\infty} \rangle & = & 
\langle \psi_U | V A B V^{-1} |\psi_U \rangle \nonumber \\
&=& \langle \psi_U | \tilde{A} \tilde{B} |\psi_U \rangle
\eea
As a consequence any correlation function of operators involving rotated electrons
at finite $U$ may be obtained from the correlation function of the original electrons
calculated at very large $U$. The form of the unitary transformation, $V$, is rather
involved and contains infinite terms if expressed in terms of the original electron operators. However, to calculate
their correlation functions it is enough to calculate the correlation function
of the original electrons at large $U$, which greatly simplifies the problem.
(As a consequence, the correlation function of $\tilde{A} \tilde{B}$ is independent of
$U$.) It is the purpose of this work to calculate the real-space correlation functions
of these objects.

In this paper we use several methods to determine these correlation
functions such as a mean-field theory based on the Zou-Anderson transformation \cite{Zou_1988},
a possible description in terms of an exact spin-charge-like separation of
the original degrees of freedom introduced by \"Ostlund-Granath \cite{Ostlund_2006}
and the density matrix renormalization group (DMRG) technique.

The first issue considered in the following has to do with the definition of which 
correlations functions one wants to calculate. Those typically involve products
of several electronic operators. Considering low energies, where a bosonization
approach should apply, we expect that as the number of fields increases the
absolute value of the correlation functions exponents should increase (considerably).
In the metallic phase bosonization predicts a power-law decay with distance.
(If the exponent is high then the extent of the correlation function should
be very small). In the case of half-filling Umklapp scattering may change the
behavior to an exponential decay. (However, if the exponent of the power-law
decay is high the two behaviors will be to some extent similar).
One of the aims of this work is to determine the exponents of these 
decays or their correlation lengths.

\section{Correlation functions}

The correlation function,
\be
C_1(r) = \langle \left(1-n_{\uparrow}(r) \right) \left( 1-n_{\downarrow}(r) \right)
n_{\uparrow}(r=0) n_{\downarrow}(r=0)
\ee
and its connected function,
\be
C_1^c(r) = C_1(r)-\langle \left(1-n_{\uparrow}(r) \right) \left( 1-n_{\downarrow}(r) \right) \rangle
\langle n_{\uparrow}(r=0) n_{\downarrow}(r=0) \rangle
\ee
contain information about the correlations between a unoccupied site (of the
original electrons) at point $r$ and a doubly occupied site at the origin. This is
a charge correlation function. It is related (at large $U$) with the proposed
(anti-)bound states of $\eta$-spinons with opposite $\eta$-spin projections.
Those refer to the $\eta$-spin degrees of freedom of pairs of rotated-electron
doubly occupied and unoccupied sites.

The correlation function,
\be
C_2(r) = \langle \left(1-n_{\uparrow}(r) \right) n_{\downarrow}(r)
n_{\uparrow}(r=0) \left( 1-n_{\downarrow}(r=0) \right) \rangle
\ee
and its connected function,
\be
C_2^c(r) = C_2(r)-\langle \left(1-n_{\uparrow}(r) \right) n_{\downarrow}(r) \rangle
\langle n_{\uparrow}(r=0) \left( 1-n_{\downarrow}(r=0) \right) \rangle
\ee
are related (at large $U$) with the proposed bound states of spinons with
opposite spin projections. This is a spin-like correlation function.
Even though the connection to the Bethe-ansatz states is through the rotated
electrons, and only in the large $U$ limit they are close to the original
electrons, we will consider in this work the correlation functions at different
values of $U$.

We also calculate some mixed correlations where at point $r$ we have for instance a doubly
occupied site and at site $r=0$ a singly occupied site, or at site $r$ a unoccupied
site and at site $r=0$ a singly occupied site. That is we also calculate, for instance,
\be
C_3(r) = \langle \left(1-n_{\uparrow}(r) \right) \left( 1-n_{\downarrow}(r) \right)
n_{\uparrow}(r=0) \left( 1-n_{\downarrow}(r=0) \right) \rangle
\ee
or
\be
C_4(r) = \langle n_{\uparrow}(r) n_{\downarrow}(r)
n_{\uparrow}(r=0) \left( 1-n_{\downarrow}(r=0) \right) \rangle
\ee
and the corresponding connected correlation functions.

The operator that counts the number of doubly occupied sites may be written
as $\sum_r c_{r,\uparrow}^{\dagger} c_{r,\uparrow} c_{r,\downarrow}^{\dagger} c_{r,\downarrow}$,
and similarly for the unoccupied sites 
$\sum_r c_{r,\uparrow} c_{r,\uparrow}^{\dagger} c_{r,\downarrow} c_{r,\downarrow}^{\dagger}$,
singly occupied sites with spin up
$\sum_r c_{r,\uparrow}^{\dagger} c_{r,\downarrow} c_{r,\downarrow}^{\dagger} c_{r,\uparrow}$ and
singly occupied sites with spin down
$\sum_r c_{r,\downarrow}^{\dagger} c_{r,\uparrow} c_{r,\uparrow}^{\dagger} c_{r,\downarrow}$.
The evaluation of the correlation functions depends on the band-filling. Away from half-filling
(metallic phase) we expect from the bosonisation that both the charge and the spin
correlation functions will decay with distance, as power laws. At half filling it is expected
that the charge correlation functions become exponential-like, due to the presence of
a charge gap. In finite magnetic field the spin degrees of freedom will also develop a gap.

The corresponding correlation functions are then expected to be of the form,
\be
C(r) \sim \frac{1}{r^{\sigma}} e^{-\frac{r}{\xi}}
\label{exponential}
\ee
for the charge correlation functions, if there is a charge gap such as at half-filling, 
where $\xi$ is the correlation length, and with a possible extra oscillating factor of the type $(-1)^r$. 
On the other hand, the spin correlation functions and the charge correlation
functions in the metallic phase are expected to be of the form,
\be
C(r) \sim \frac{1}{r^{\alpha}} [ \ln r]^{\beta}
\label{power}
\ee
also with a possible extra oscillating factor of the type $(-1)^r$. 

In the half-filling case and in the limit of large $U$ the spin part of the
Hubbard model reduces to the spin-$1/2$ isotropic Heisenberg chain. Also the charge part is gapped.
Previous studies for the charge correlations suggest that $\sigma \sim 1/2$. The study of 
the spin-spin correlation functions at half-filling lead to some controversy about the presence of
logarithmic corrections, but the presence of a logarithmic factor with exponent $\beta=0.5$ was
confirmed and the above $\alpha$ exponent reading $\alpha \sim 1$ \cite{Affleck,Giamarchi,Singh,Hallberg,Hikihara}.

There are transformations proposed in the literature that lead to
a similar decoupling of the electronic degrees of freedom.
Examples are for instance given in \cite{Zou_1988} or in \cite{kotliar}.
The main motivation was the study of either the large-$U$ limit in
the Hubbard or Anderson models \cite{dorin}, with the intent of controlling
in a efficient way the projection to states where double occupancy
is restricted (as in the $t-J$ model), but considering a finite
value of $U$ instead of the extreme case of infinite $U$, usually
taken care of by a single slave boson \cite{coleman}. 
Both representations introduce explicitly operators related to the four possible
states associated with each site. Namely that a site may be unoccupied, singly
ocuppied with a given spin projection or douby occupied. In the Kotliar and
Ruckenstein procedure four bosonic operators are added, enlarging the operator
space, that act as projectors on the original fermionic operators. In the Zou-Anderson
transformation the original electron operators are replaced by two sets of two 
bosonic and fermionic operators that fulfill the projection. However, both representations lead
to an enlargement of the physical Hilbert space and the extra unphysical states have to
be projected out. We note, however, that
the representation introduced by Zou and Anderson (ZA) has been used to explicitly
obtain an exact solution of the Hubbard model in the large $U$ limit
in a much simpler way as compared to the Bethe ansatz \cite{ricardo}.
Also, it has been used to study the stiffness of the one-dimensional
Hubbard model in a way equivalent and alternative to the Bethe-ansatz solution \cite{stiffness}.

\subsection{Zou-Anderson transformation}

The electron operators may be written as,
\bea
c_{i,\sigma} &=& e_i^{\dagger} S_{i,\sigma} + \sigma S_{i,-\sigma}^{\dagger} d_i \nonumber
\\
c_{i,\sigma}^{\dagger} &=& S_{i,\sigma}^{\dagger} e_i + \sigma d_i^{\dagger} S_{i,-\sigma}
\eea
where the operators $e_i, d_i, S_{i,\sigma}$ annihilate sites that are unoccupied,
doubly-occupied and singly-occupied with an electron with spin $\sigma$, respectively.
Since the electron operators are fermionic we can either choose the operators
$e,d$ as bosonic and the operators $S_{i,\sigma}$ as fermionic, or vice-versa.
In the original paper \cite{Zou_1988} the first choice was made (called slave-boson approach) and
in \cite{ricardo} the second choice was taken (called slave-fermion
approach). Most expressions are the same,
formally, in either case. The difference arises when one integrates over degrees
of freedom or in the mean-field approach when the Bose-Einstein or the Fermi-Dirac
distributions appear.

The enlargement of the degrees of freedom imposes the constraint,
\be
e_i^{\dagger} e_i + d_i^{\dagger} d_i + \sum_{\sigma} S_{i,\sigma}^{\dagger} S_{i,\sigma}
=1
\ee
at each site. (This is the completeness relation of the four possibilities: one site
is either unoccupied, doubly-occupied, or is singly occupied by an electron with spin-up or
down). The constraint is simply obtained imposing the anticommutation
relation $\{ c_{i,\sigma},c_{i,\sigma} \} =1$ and considering either the slave-bosons
or slave-fermions commutation or anticommutation relations.

Let us consider the Hubbard model written as,
\be
H = -t \sum_{i,\delta; \sigma} c_{i,\sigma}^{\dagger} c_{i+\delta,\sigma}
+U \sum_i n_{i,\uparrow} n_{i,\downarrow} - \mu \sum_{i,\sigma} c_{i,\sigma}^{\dagger}
c_{i,\sigma}
\ee
The first is the hopping term between a site $i$ and its neighbors distant by $\delta$
(in general vectors in a $d$-dimensional space), $U$ is the on-site repulsion
and $\mu$ the chemical potential enforcing the band filling.

In terms of the slave-bosons or slave-fermions the Hubbard Hamiltonian may be rewritten as,
\bea
H &=& -t \sum_{i,\delta;\sigma} \left( e_i e_{i+\delta}^{\dagger}
-d_i d_{i+\delta}^{\dagger} \right) S_{i,\sigma}^{\dagger} S_{i+\delta,\sigma}
\nonumber \\
&-& t \sum_{i,\delta} \left( e_i d_{i+\delta} (S_{i\uparrow}^{\dagger}
S_{i+\delta,\downarrow}^{\dagger} - S_{i\downarrow}^{\dagger} 
S_{i+\delta,\uparrow}^{\dagger} ) +
d_i^{\dagger} e_{i+\delta}^{\dagger} (S_{i\downarrow}
S_{i+\delta,\uparrow}- S_{i\uparrow} 
S_{i+\delta,\downarrow} ) \right) \nonumber \\
&+& U \sum_i d_i^{\dagger} d_i + \mu \sum_i \left( e_i^{\dagger} e_i - d_i^{\dagger}
d_i \right) - \mu N_a
\eea

The ZA mapping reverses the role of the interacting and kinetic terms
in the Hamiltonian. The interacting Hubbard term becomes quadratic
in the ZA particles and the kinetic one is transformed into an interacting
quartic term that couples particles along the lattice links.
This is particularly useful to study the strongly interacting (large $U$) regime
where the kinetic term is treated as a perturbation. The price of
this transformation is the appearance of an on-site
constraint, which assures exactly one particle per lattice site. In the mean field (MF)
approach this translates to an on-site Lagrange multiplier.

The problem to be solved involves the effective Hamiltonian,
\bea
H &=& -t \sum_{i,\delta;\sigma} \left( e_i e_{i+\delta}^{\dagger}
-d_i d_{i+\delta}^{\dagger} \right) S_{i,\sigma}^{\dagger} S_{i+\delta,\sigma}
\nonumber \\
&-& t \sum_{i,\delta} \left( e_i d_{i+\delta} (S_{i\uparrow}^{\dagger}
S_{i+\delta,\downarrow}^{\dagger} - S_{i\downarrow}^{\dagger} 
S_{i+\delta,\uparrow}^{\dagger} ) +
d_i^{\dagger} e_{i+\delta}^{\dagger} (S_{i\downarrow}
S_{i+\delta,\uparrow}- S_{i\uparrow} 
S_{i+\delta,\downarrow} ) \right) \nonumber \\
&+& U \sum_i d_i^{\dagger} d_i + \mu \sum_i \left( e_i^{\dagger} e_i - d_i^{\dagger}
d_i \right) - \mu N \nonumber \\
&+& \sum_i \lambda_i \left( e_i^{\dagger} e_i + d_i^{\dagger} d_i + \sum_{\sigma} S_{i,\sigma}^{\dagger}
S_{i,\sigma} -1 \right)
\eea
where we have introduced at each site a Lagrange multiplier, $\lambda_i$, to inforce the constraint.
The transformation of the electron operators to the auxiliary operators already
embodies part of the classification of the Bethe-ansatz states (actually for the rotated
electrons). It seems therefore natural to decouple the quartic terms in the Hamiltonian
in such a way that the unoccupied and doubly-occupied sites are separated, on a first stage,
from the singly-occupied sites. Also, we consider that the unoccupied sites and the
doubly-occupied sites are paired on nearest-neighbor links. On the other hand, the spin states of the singly-occupied
sites are paired into spin singlets. 

We then consider the mean-field Hamiltonian in the following form \cite{u1paper},
\bea
 H_{MF} = 
- t \sum_{i,\delta;\sigma} \left\{ \left( \chi_{\delta}^e - \chi_{\delta}^d \right) 
S_{i,\sigma}^{\dagger} S_{i+\delta,\sigma} \right. \nonumber \\
+ \left. \left( e_{i+\delta}^{\dagger} e_i - d_{i+\delta}^{\dagger} d_i \right) \chi_{\delta,\sigma}^S
- \left( \chi_{\delta}^e - \chi_{\delta}^d \right) \chi_{\delta, \sigma}^{S} \right\} \nonumber \\
- t \sum_{i,\delta} \left\{ \Phi_{\delta} \left( S_{i\uparrow}^{\dagger}
S_{i+\delta,\downarrow}^{\dagger} - S_{i\downarrow}^{\dagger} 
S_{i+\delta,\uparrow}^{\dagger} \right) +
\Phi_{\delta}^* \left( S_{i\downarrow}
S_{i+\delta,\uparrow}- S_{i\uparrow} 
S_{i+\delta,\downarrow} \right) \right. \nonumber \\ 
\left. +e_i d_{i+\delta} \Delta_{\delta}^* + d_i^{\dagger} e_{i+\delta}^{\dagger} \Delta_{\delta}
-\Phi_{\delta} \Delta_{\delta}^* - \Phi_{\delta}^* \Delta_{\delta} \right\} 
\nonumber \\
+ U \sum_i d_i^{\dagger} d_i + \mu \sum_i \left( e_i^{\dagger} e_i - d_i^{\dagger}
d_i \right) - \mu N \nonumber \\
+ \sum_i \lambda_i \left( e_i^{\dagger} e_i + d_i^{\dagger} d_i + \sum_{\sigma} S_{i,\sigma}^{\dagger}
S_{i,\sigma} -1 \right)
\label{mftH}
\eea
The quantities appearing in this Hamiltonian expression are defined as follows,
\bea
\chi_{\delta}^e &=& \langle e_{i+\delta}^{\dagger} e_i \rangle \nonumber \\
\chi_{\delta}^d &=& \langle d_{i+\delta}^{\dagger} d_i \rangle \nonumber \\
\chi_{\delta,\sigma}^S &=& \langle S_{i,\sigma}^{\dagger} S_{i+\delta,\sigma} \rangle \nonumber \\
\Phi_{\delta} &=& \langle e_{i} d_{i+\delta} \rangle \nonumber \\
\Delta_{\delta} &=& \langle \left(S_{i,\downarrow} S_{i+\delta, \uparrow}
-S_{i, \uparrow} S_{i+\delta, \downarrow} \right) \rangle 
\eea
Besides considering hopping amplitudes we also introduce two pairing terms \cite{u1paper}, one between the
unoccupied and doubly-occupied sites and another one between singly occupied sites with
opposite spins. Note that one refers to a boson pairing and the other to a fermionic pairing. 
The choice of the mean-field parameters has in mind the possible
bound states between the $e$ and $d$ operators and the $S_{\uparrow}$ and
$S_{\downarrow}$ operators. Indeed, we intend to investigate the tendency to
form these bound-states. 
The problem is now quadratic and may be diagonalized. The solution is briefly 
reviewed in Appendix A.

Generically the phases found by solving the MF solutions for arbitrary band filling
and energy are characterized as follows \cite{u1paper}:
Phase (1) is conducting and characterized by $\chi\neq 0,\ \Delta=0$. In it
the spinons are gapless and the charge degrees of freedom exibit
a gap of the order of the temperature, which closes at $T=0$.
This is the lowest free-energy phase. Within the mean-field approach
it is that corresponding to the ground state.
At finite energies other phases emerge \cite{u1paper}, such as
phase (2) $\chi=0,\ \Delta\neq0$, which is gapped for both
degrees of freedom. Since it appears near $x=0$ it is tempting to
identify it with an insulating antiferromagnet.
Phase (3) $\chi\neq0,\ \Delta\neq0$ is a precursor of
the superconductor. In it there exists spin-singlet formation
but the charge motion is incoherent since no condensation is allowed.
If one imposes $e_{k=0}=Z$, this phase splits into two
sub-phases, analog to the pseudogap and superconducting phases in \cite{Lee_2006}.
Phase (4) is an incoherent high-temperature phase where all correlations are zero.

The hopping and pairing correlation functions between two sites at distance $r$ from each
other are given by,
\begin{eqnarray}
\chi_{F}(r) & = & \av{\mathit{s}_{r,1}^{\dagger}\mathit{s}_{0,1}+\mathit{s}_{r,-1}^{\dagger}s_{0,-1}}_{0},\nonumber \\
\Delta_{F}(r) & = & \av{\mathit{s}_{r,1}\mathit{s}_{0,-1}-\mathit{s}_{r,-1}\mathit{s}_{0,1}}_{0},\nonumber \\
\chi_{B}(r) & = & \av{\mathit{d}_{r}^{\dagger}\mathit{d}_{0}-\mathit{e}_{r}^{\dagger}\mathit{e}_{0}}_{0},\nonumber \\
\Delta_{B}(r) & = & \av{\mathit{d}_{r}\mathit{e}_{0}+\mathit{e}_{r}\mathit{d}_{0}}_{0}.\label{eq:non_phys_corr}
\end{eqnarray}
They have been calculated before in Ref. \cite{u1paper} for the various phases.
Although these correlation functions are not gauge invariant, they are
useful to characterize the different phases. For phases (1) and (3) and
both for the fermion and the boson hopping correlation
functions, it was found close to half filling that the correlation length increases as the doping increases \cite{u1paper}.
Particularly, the bosonic correlation function has a large correlation
length. Analyzing the correlation length of $\Delta_{B}$, one clearly
sees a long-range correlation in the high doping regime ($x=1-n$ large), possibly precursor
of Bose-condensation and superconductivity. In the low-doping region,
both the bosonic and the fermionic correlation functions have a smaller
range consistent with a spin gapped state. In this regime the two
correlation functions have similar range, while at higher doping the
charge correlation function has a much larger range compared to the
spin correlation function.
These correlation functions will be relevant if the system is in a deconfined
phase. In a confined phase these correlation functions loose their significance, as
the various degrees of freedom are confined within the real electrons. However,
use of the Bethe-ansatz solution reveals that some fractionalization and rearrangement
of the degrees of freedom occurs. 

We can now calculate the above mentioned correlation functions, $C_1(r), C_2(r)$, in the
mean-field approach.

\begin{figure}[t]
\begin{centering}
\includegraphics[width=0.49\columnwidth]{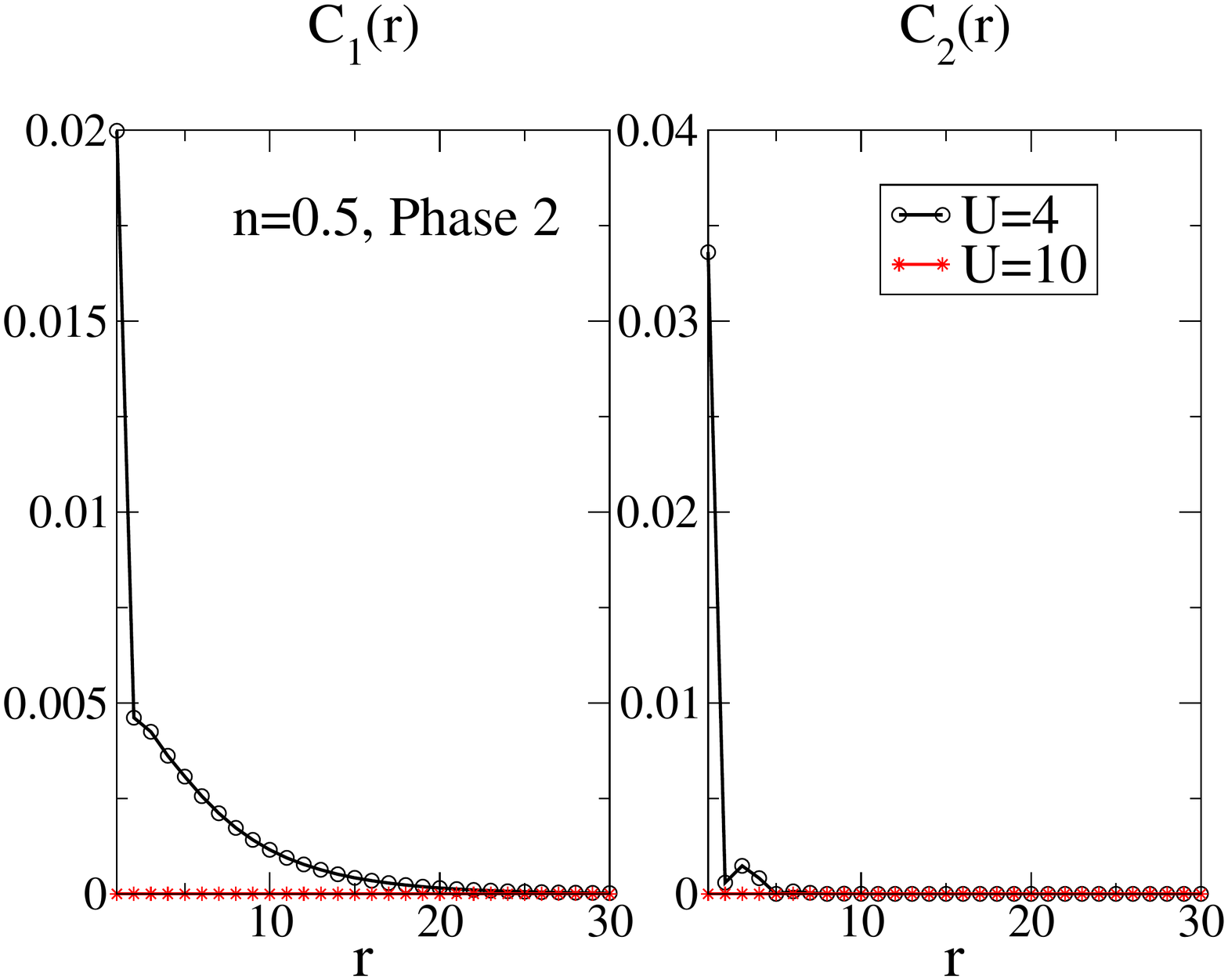}
\includegraphics[width=0.49\columnwidth]{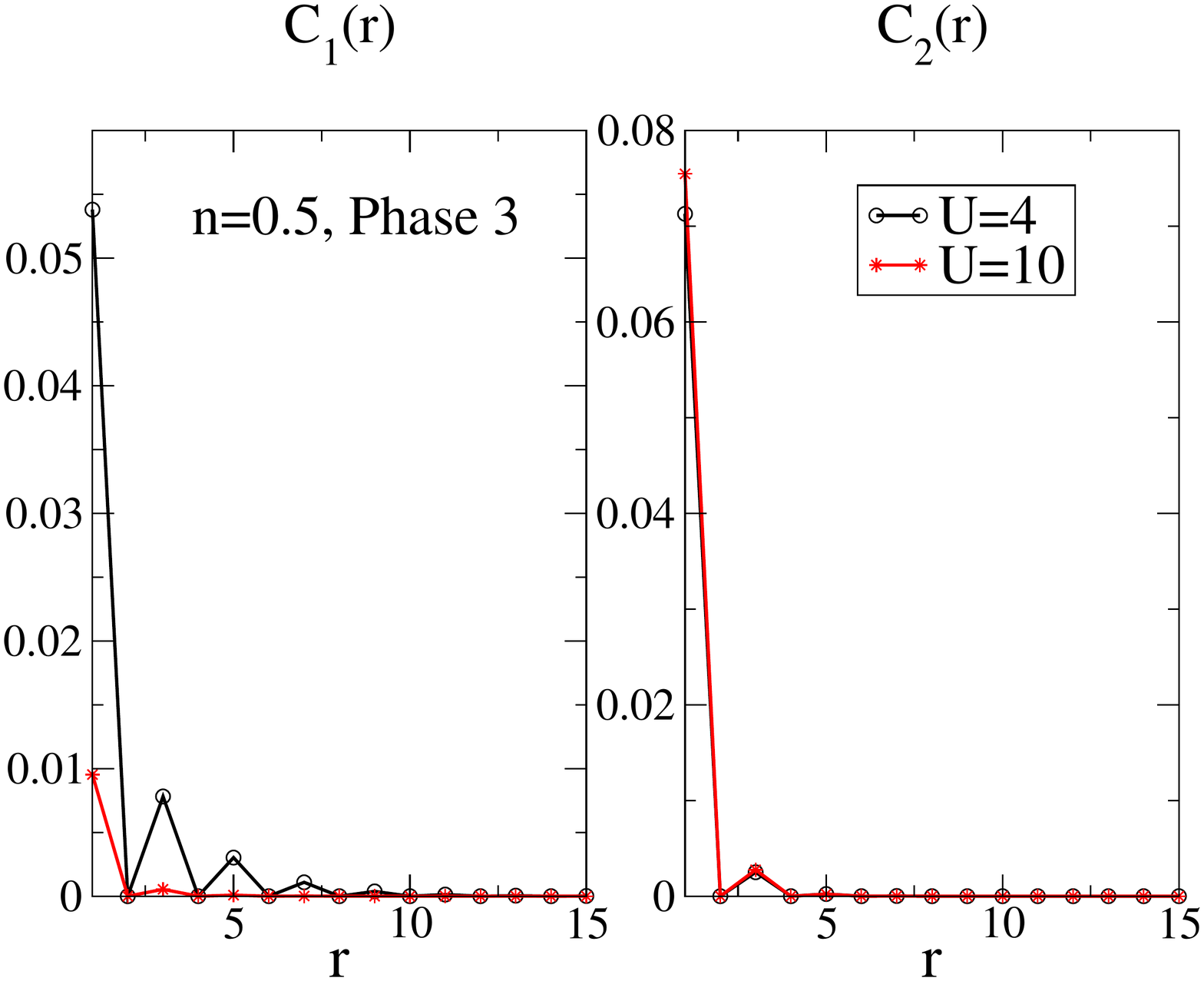}
\par\end{centering}
\caption{\label{mfields2} Correlation functions $C_1^c(r), C_2^c(r)$
for $n=0.5$, $U=4,10$ and $N=100$ in the phases 2 and 3.}
\end{figure}
\begin{figure}[t]
\begin{centering}
\includegraphics[width=0.49\columnwidth]{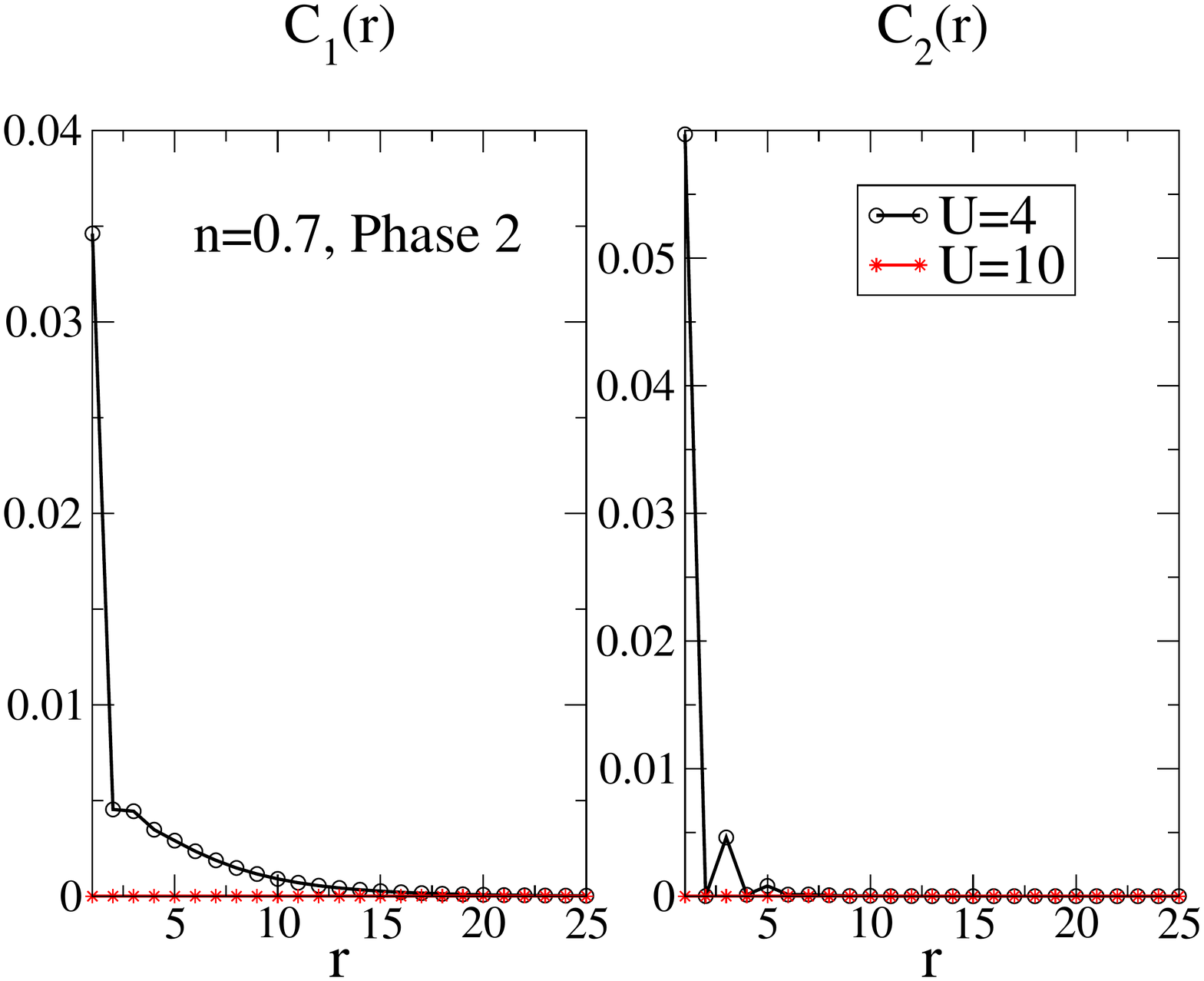}
\includegraphics[width=0.49\columnwidth]{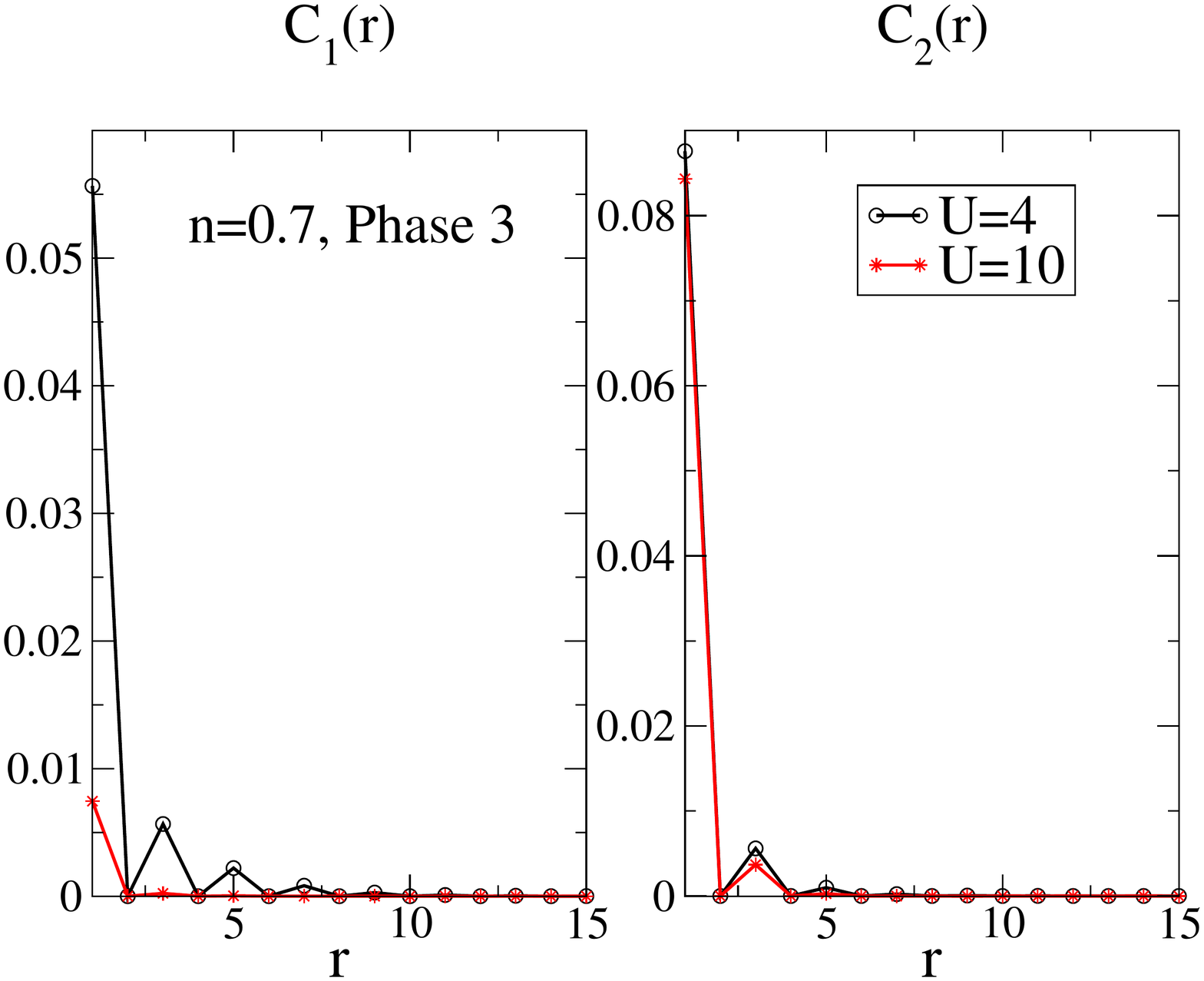}
\par\end{centering}
\caption{\label{mfields3} Correlation functions $C_1^c(r), C_2^c(r)$
for $n=0.7$, $U=4,10$ and $N=100$ in the phases 2 and 3.
}
\end{figure}
\begin{figure}[t]
\begin{centering}
\includegraphics[width=0.49\columnwidth]{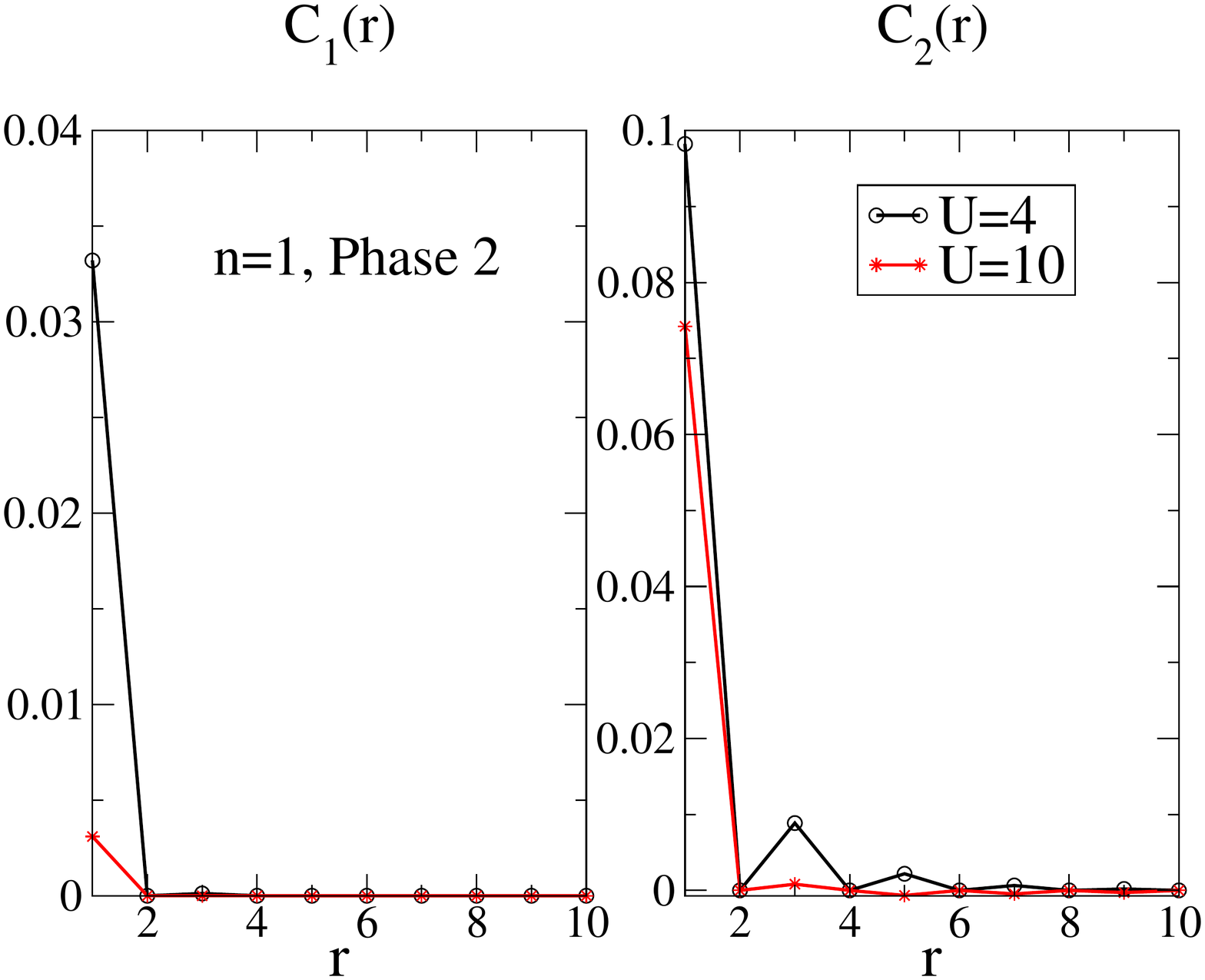}
\includegraphics[width=0.49\columnwidth]{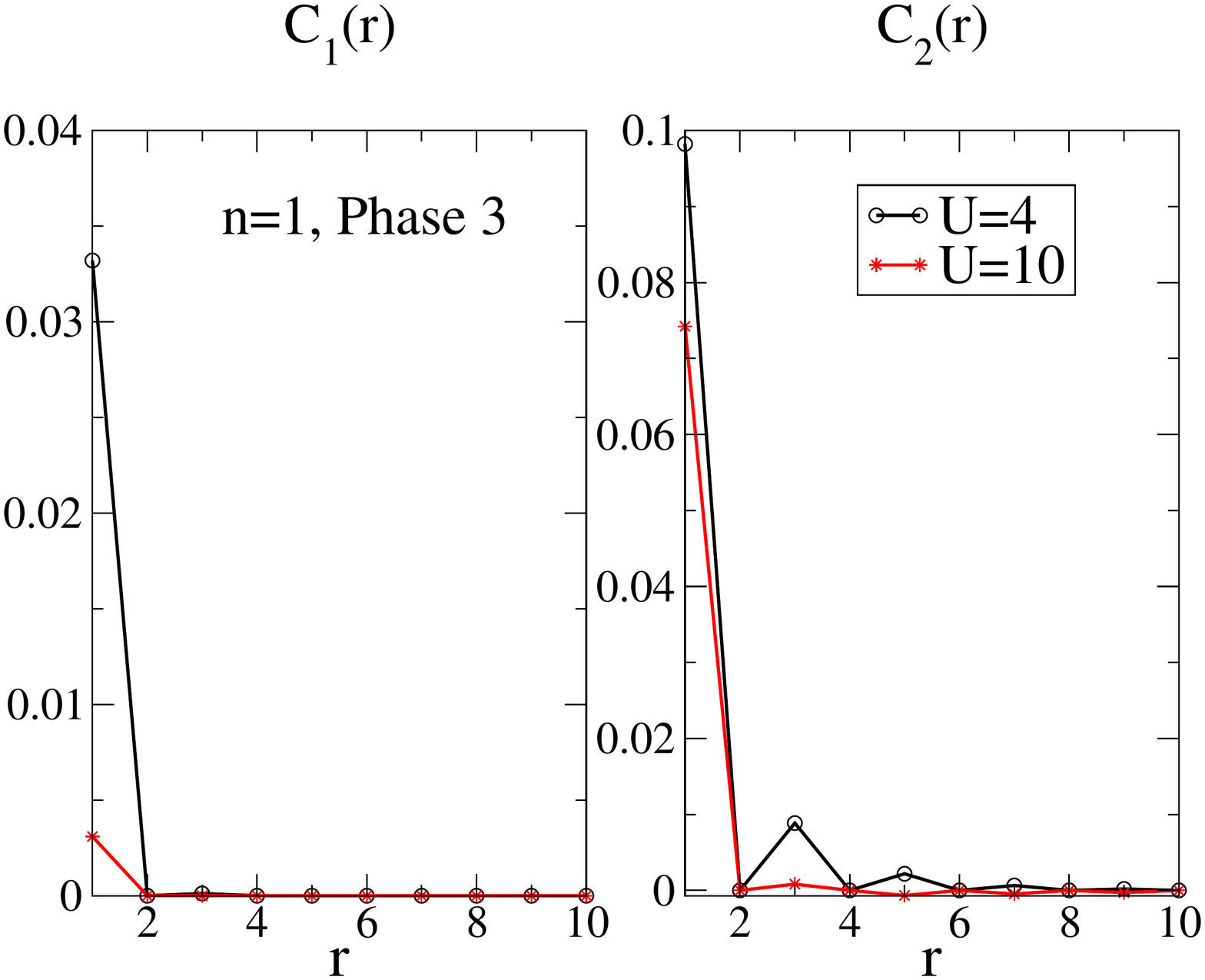}
\par\end{centering}
\caption{\label{mfields1} Correlation functions $C_1^c(r), C_2^c(r)$
for $n=1$, $U=4,10$ and $N=100$ in the phases 2 and 3.
}
\end{figure}

Using the constraint we can write that,
\bea
C_1(r) &=& \langle \left(1-n_{\uparrow}(r) \right) \left( 1-n_{\downarrow}(r) \right)
n_{\uparrow}(r=0) n_{\downarrow}(r=0) \rangle  = \langle n_e(r) n_d(0) \rangle \nonumber \\
C_2(r) &=& \langle \left(1-n_{\uparrow}(r) \right) n_{\downarrow}(r) 
n_{\uparrow}(r=0) \left( 1-n_{\downarrow}(r=0) \right) \rangle = \langle n_{s \downarrow}(r)
n_{s \uparrow}(0) \rangle \nonumber \\
& & 
\eea
and the corresponding connected functions read,
\bea
C_1^c(r) &=& \langle n_e(r) n_d(0) \rangle -\langle n_e(r) \rangle \langle n_d(0) \rangle
\nonumber \\ 
C_2^c(r) &=& \langle n_{s \downarrow}(r) n_{s \uparrow}(0) \rangle -
\langle n_{s \downarrow}(r) \rangle \langle n_{s \uparrow}(0) \rangle
\eea
where $n_e= e^{\dagger} e$, $n_d=d^{\dagger} d$, $n_{s \uparrow}= 
s_{\uparrow}^{\dagger} s_{\uparrow}$ and $n_{s \downarrow}= 
s_{\downarrow}^{\dagger} s_{\downarrow}$. Using their representations in
momentum space in terms of the diagonalized operators we obtain the results
derived in Appendix A. Those are shown in Figs. \ref{mfields2} and \ref{mfields3}. 

At half filling we are in the insulating phase (2). We expect therefore a charge
gap and an exponential decay of the correlation function $C_1(r)$. This is indeed
seen for the values of $U=4,10$, and independentely of the system size, where the 
spatial extent refers basically to nearest-neighbors. The spin part is gapless, so that 
a larger range correlation function is expected, as shown in Fig. \ref{mfields1}. As $U$ increases,
the magnitude of $C_2(r)$ grows for $r=1$, but decreases faster with distance as compared
to smaller values of $U$. 
Away from half filling we consider the densities $n=0.7,0.5$. In the metallic phase the charge correlation
function has a much larger range, comparable to or larger than the spin counterpart.
This qualitatively agrees with the results for the non-gauge invariant correlations.
Far from half filling (quarter filling, $n=0.5$) the spin correlations decrease fast with
distance since we move far from the half-filled antiferromagnet.

As stated above, there are two sorts of approximations within the present approach. The first
is related to the enlargment of the physical Hilbert space and the necessity of introducing a constraint,
to reduce the system to that space. The other sort of approximation has to do
with the mean-field approach used.
Moreover, the constraint is only implemented on average, as usual in slave-boson or slave-fermion approaches.

The first difficulty has been overcome recently \cite{Ostlund_2006}, with the introduction of an exact transformation
of the electron operators in terms of other operators that are related to the spin-charge
separation of the model. The electron operators can be written as composites of charge-like and spin-like
operators that do not give rise to any unphysical states, and thus avoids introducing any constraint.

\subsection{\"Ostlund-Granath transformation}

This transformation \cite{Ostlund_2006} introduces new operators called
quasicharge $ \chop{r} $ and quasispin operators $ \gee{i}{r} $
that obey, respectively, Fermi and Bose statistics,
\begin{eqnarray} \label{eq:toquasi}
\chop{r} & = &  \cdop{\uparrow}{r}( 1 - n_{\downarrow,\,r}) + (-1)^r \cop{\uparrow}{r} n_{\downarrow , \, r} \\ 
\gee{+}{r} & = & ( \cdop{\uparrow}{r} - ( -1)^r \cop{\uparrow}{r} \; ) \; \cop{\downarrow}{r} \\ \nonumber
\gee{-}{r} & = &  (\gee{+}{r})^{\dagger} \\ \nonumber
\gee{z}{r}  & = &  \textstyle{\half}  -  n_{\downarrow,\,r} 
\end{eqnarray}
These operators satisfy the algebra
$ \{ \chop{r}, \chdop{\rp} \} =  \delta_{r,\rp}$ ,
$ \{ \chdop{r}, \chdop{\rp} \} = 0 $ ,
$ \left[\, \chdop{r}, \gee{i}{\rp} \, \right] = 0 $ ,
$ \left[ \, \gee{i}{r}, \gee{j}{\rp} \right] = i \delta_{r\rp}  \sum_{k} \epsilon_{ijk} \gee{k}{r} $.
The electron operators are expressed in terms of them as,
\begin{eqnarray}
\cdop{\uparrow}{r} & =  & \chop{r} \; ( \, \half + \gee{z}{r} ) \;+ 
        \; (-1)^r \; \chdop{r} (\half - \gee{z}{r} ) \\ \nonumber
\cdop{\downarrow}{r}   & =  & \gee{-}{r} \, ( \,  \chop{r} \; - 
        \; (-1)^r \,  \chdop{r} \,  ).
\end{eqnarray}
The \"Ostlund-Granath representation involves other operators such as
the quasicharge operator 
$ n^c_r  = \chdop{r} \chop{r} $ and
the local pseudospin operators
$ p^{i}_r = n^c_r \gee{i}{r} $, which are the generators of the $SU(2)$ algebra that corresponds to ''rotations'' between the unoccupied and
doubly occupied states \cite{shiba72}. 

\begin{figure}[t]
\begin{centering}
\includegraphics[width=0.45\columnwidth]{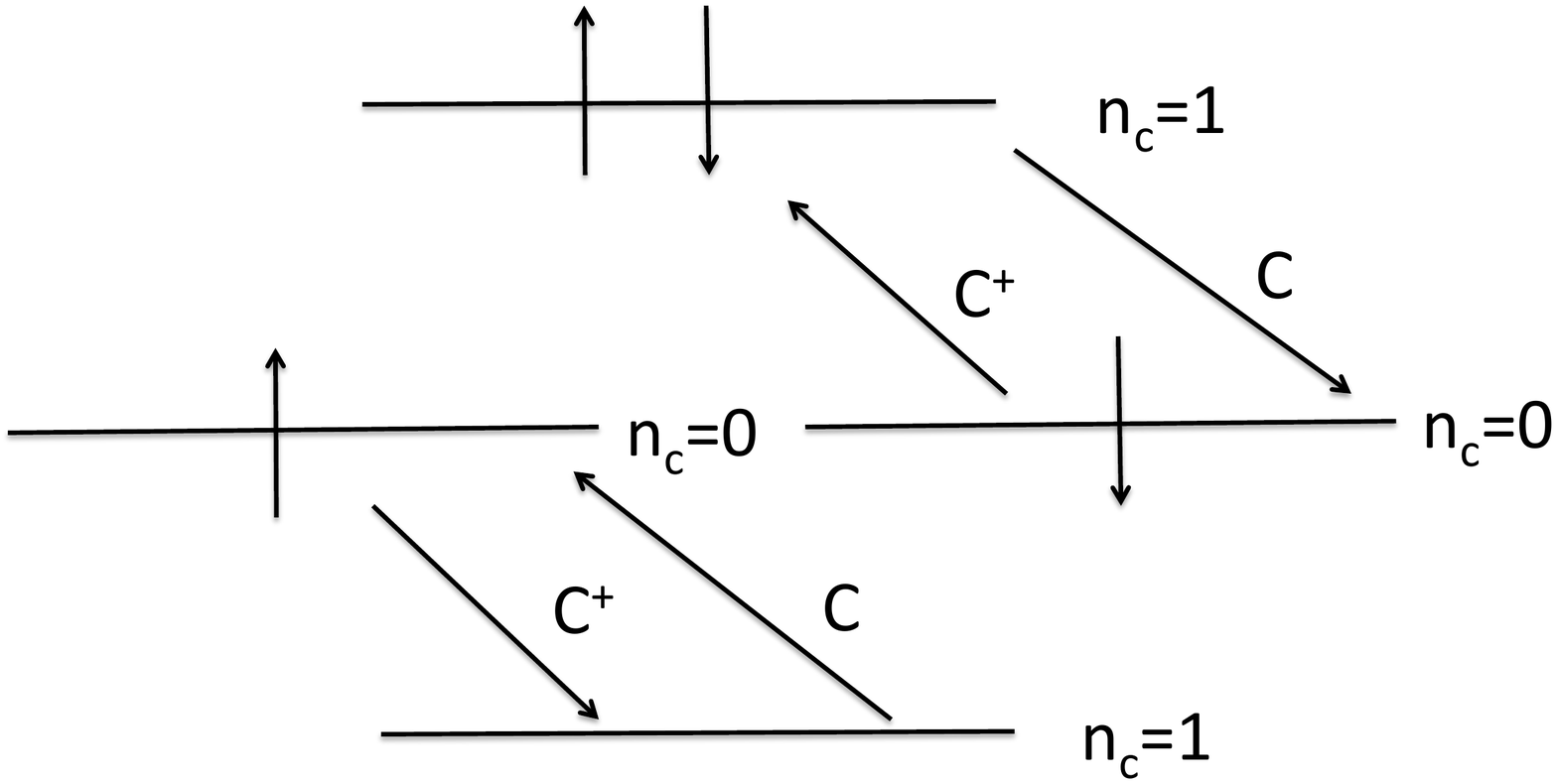}
\includegraphics[width=0.45\columnwidth]{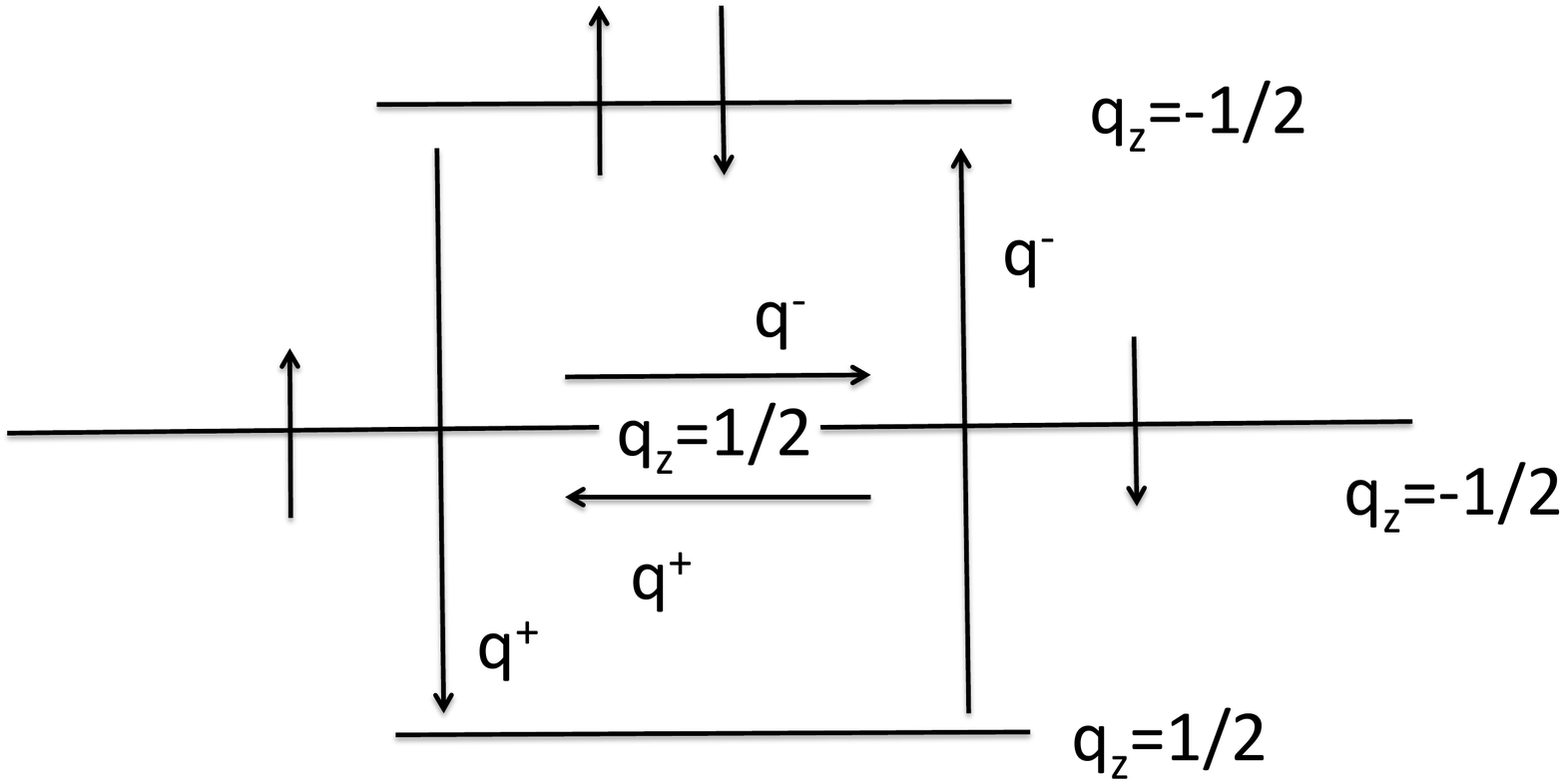}
\includegraphics[width=0.45\columnwidth]{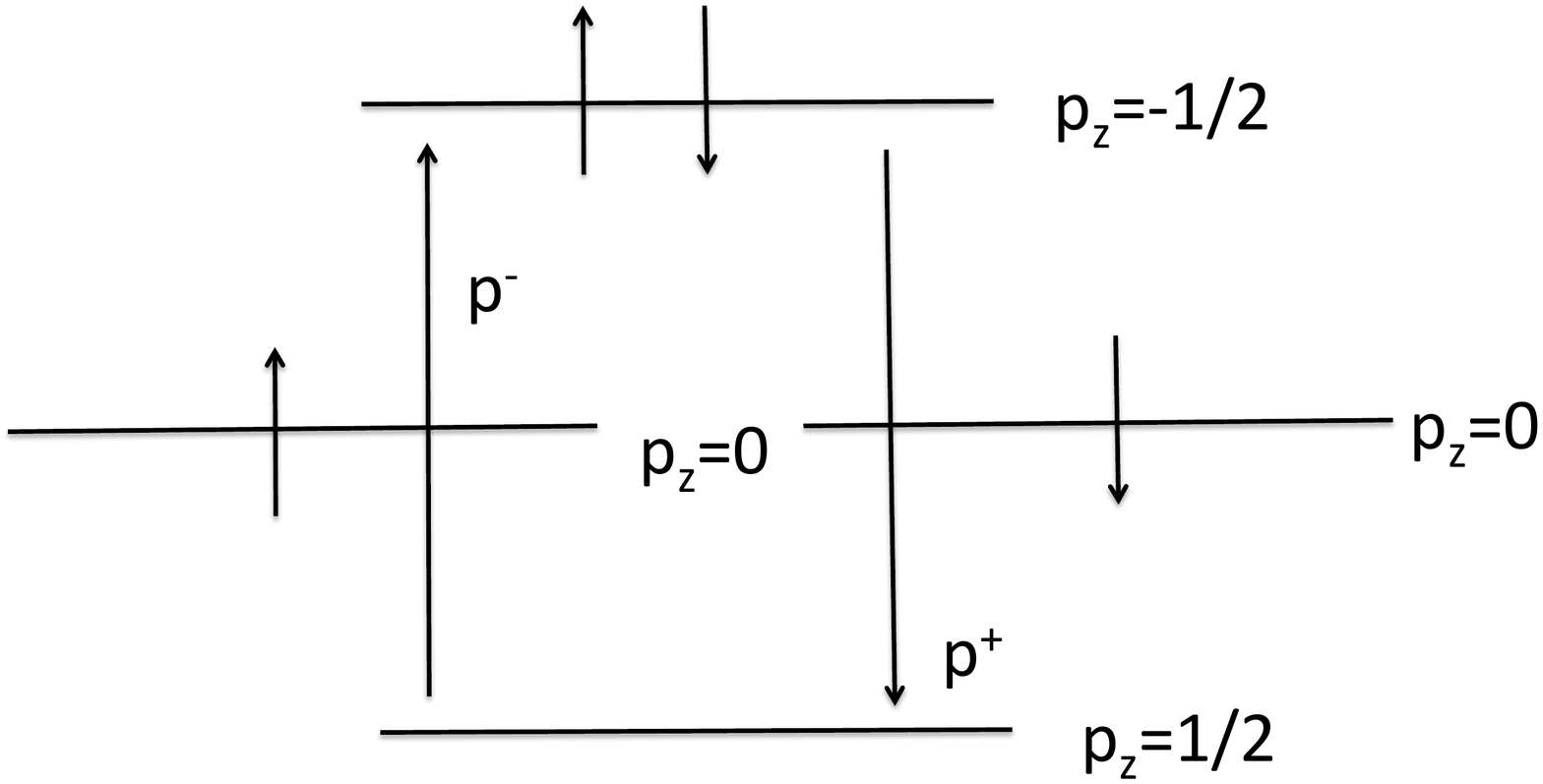}
\includegraphics[width=0.45\columnwidth]{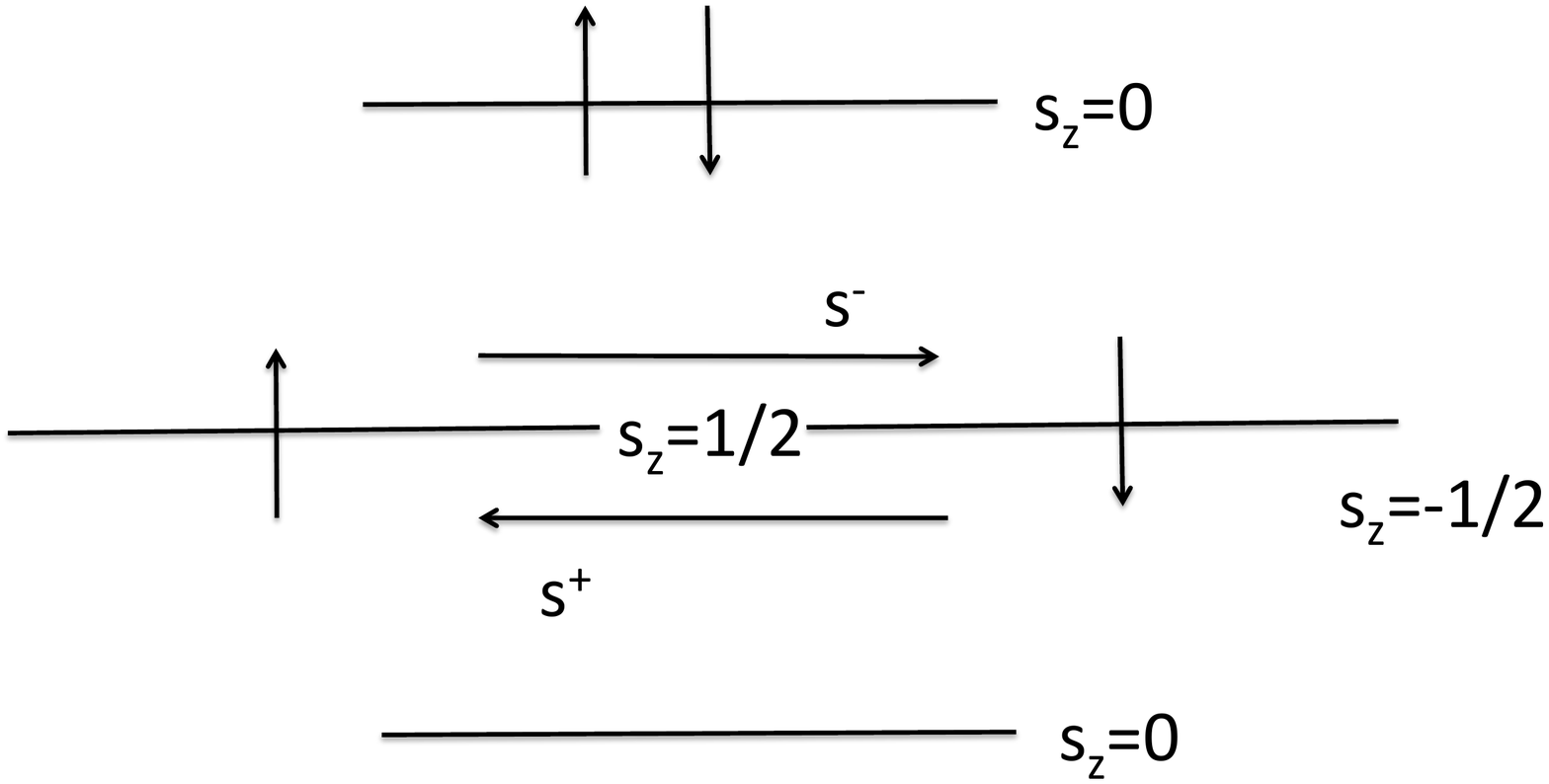}
\includegraphics[width=0.45\columnwidth]{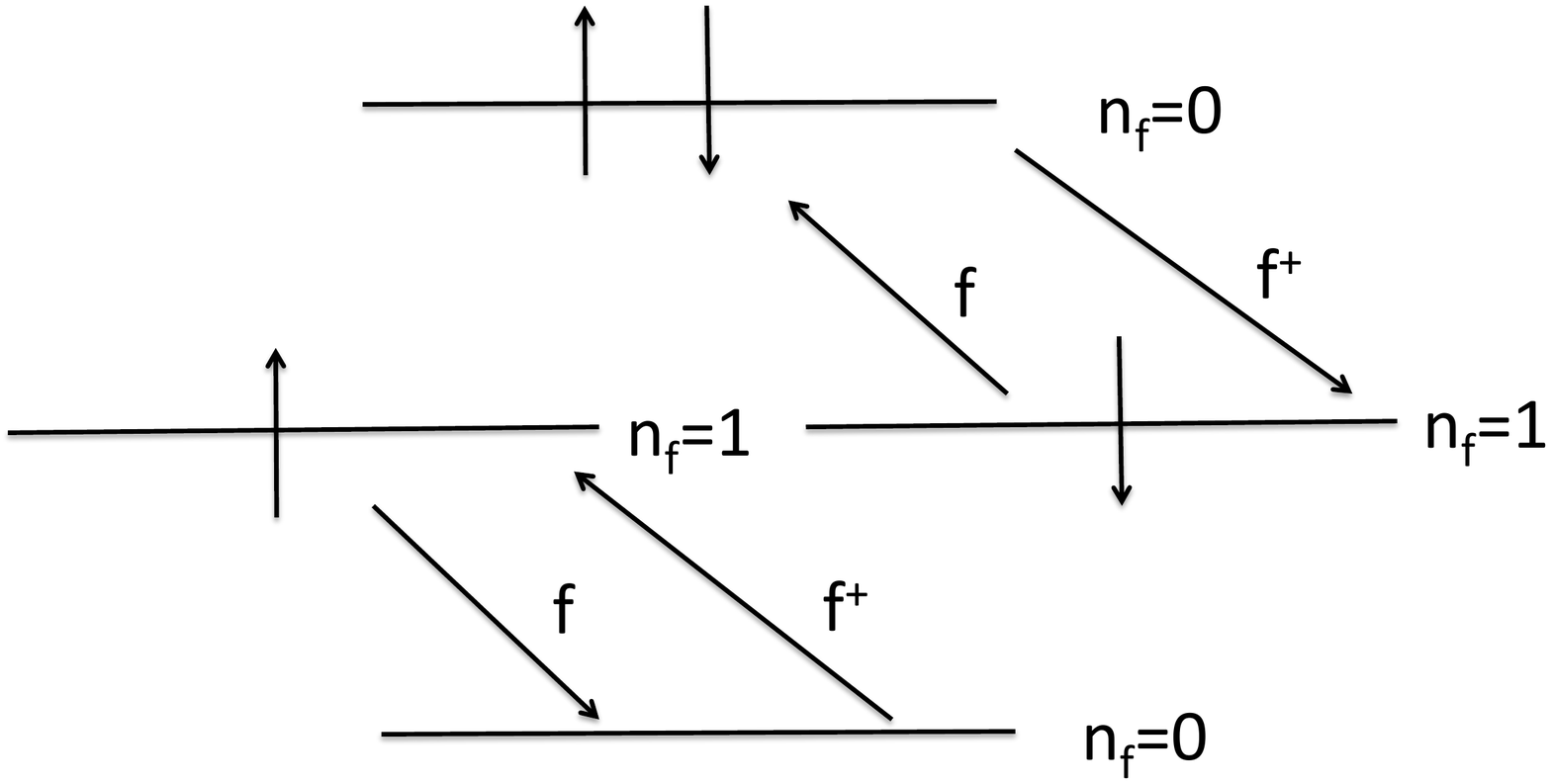}
\par\end{centering}
\caption{\label{schemes} Eigenvalues of the quasicharge, quasispin, pseudospin and spin
for each of the four basis states referring to a doubly occupied site, singly occupied site with spin up
and spin down and unoccupied site. The transitions generated by the off-diagonal operators are also shown. 
In the last panel we consider the transitions generated by $n_f=1-n_c$ where $f=c^{\dagger},f^{\dagger}=c$,
similarly to the operators introduced in \cite{Carmelo_2010} (for the rotated electrons).}
\end{figure}

The following results hold
$ n_r   =   1   - 2  n^c_r \gee{z}{r} $,
$ s^i_r  =   ( 1 - n^c_r \, ) \;  \gee{i}{r} $ and
$ n^c_r  =  (n_r - 1)^2  $
where $ s^i_r = \half \sum_{\alpha,\beta} \cdop{\alpha}{r} \sigma^i_{\alpha \beta } \cop{\beta}{r} $,
with $ \sigma^i  $ the Pauli matrices.
The total z-component of pseudospin
can therefore be seen to be half the number of doubly occupied
sites minus the number of unoccupied sites, which is precisely the charge
relative to half filling. 
The action of these operators onto the four-state basis is shown in Fig. \ref{schemes}.

The Hubbard model can be rewritten in terms of the quasiparticle operators as,
\begin{equation}
\label{eq:newhub}
H = t \;  ( \teezero + \teeplus + \teeminus ) + U h_U 
\end{equation}
with $ h_U  =  \half  \sum_r  \, \chdop{r}\chop{r}  $,
\begin{eqnarray} \label{eq:teepm}
\teezero & = & \frac{1}{2} \sum_{\e{r,\rp}}  \,  ( 1 + 4 \bgee{r}\cdot\bgee{\rp} )(\chdop{r}\chop{\rp} +CC) \\ \nonumber
\teeplus & = & \frac{1}{2} \sum_{\e{r,\rp}} (-1)^r \, ( 1 - 4 \bgee{r}\cdot\bgee{\rp} )(\chdop{r}\chdop{\rp} ) \\ \nonumber
\end{eqnarray}
and $T_{-1} = T_{1}^\dagger $.

Even though this transformation achieves some sort of exact spin-charge
separation (actually quasispin and quasicharge), the Hamiltonian
has a complicated structure. Although it has quartic interacting terms,
since they involve the quasispin operators, these are typically represented
in terms of bilinear representations of fermionic or bosonic operators.
Usually these representations enlarge the physical Hilbert space and one has
to introduce constraints. One can use a Majorana fermion representation
\cite{majoranas} but this leads to an Hamiltonian where the leading
interacting term involves six operators and, therefore, the analytical
treatment is rather complicated. One can also represent the quasispin operators
using the Jordan-Wigner transformation. The transverse terms are linear
in terms of new fermionic operators (the strings cancel out since only
nearest-neighbor hoppings are considered) but the longitudinal term
is again a bilinear in fermionic operators (which leads again to
terms with six operators). An analytical treatment
would need some approximation scheme, which is known to not yield
good results in one-dimensional systems.

It is interesting however to look at the behavior of the correlation functions
of these operators. Moreover, it has been shown that 
the quasicharge operator is associated with a recently found hidden $U(1)$
symmetry of the Hubbard model \cite{bipartite} (on any bipartite lattice). Together
with the exact Bethe-ansatz solution of the 1D problem, such a symmetry has lead to a deeper
understanding of the physics of the model, including an understanding of the
dressed scattering matrix structure \cite{npbIII}. Specifically, we are interested
in the correlation functions
for the quasicharge and the quasispin operators 
of \"Ostlund and Granath written in terms of the original electron operators 
(and the corresponding connected correlation functions),
\bea
C_5(r)=\langle n_c(r) n_c(0) \rangle &=& \langle \left( n(r)-1\right)^2 \left(n(0)-1\right)^2 \rangle \nonumber \\
C_6(r)=\langle \left(1-n_c(r) \right) \left(1-n_c(0)\right) \rangle &=& 1- \langle n_c(r)\rangle -
\langle n_c(0) \rangle + \langle n_c(r) n_c(0) \rangle \nonumber \\
C_7(r) = \langle q_z(r) q_z(0) \rangle &=& \langle \left( \frac{1}{2}-n_{\downarrow}(r) \right)
\left( \frac{1}{2}-n_{\downarrow}(0) \right) \rangle \nonumber \\
& & 
\eea
where $n(r)=n_{\uparrow}(r) + n_{\downarrow}(r)$ and $n_{\sigma}=c^{\dagger}_{\sigma} c_{\sigma}$. 

We may also rewrite the pseudospin correlation functions in terms of the
original electron operators
(and the corresponding connected correlation functions),
\bea
C_8(r)&=& \langle p_z(r) p_z(0) \rangle = \langle \left( n(r)-1\right)^2 
\left( \frac{1}{2}-n_{\downarrow}(r) \right)
\left(n(0)-1\right)^2
\left( \frac{1}{2}-n_{\downarrow}(0) \right) 
\rangle \nonumber \\
& & 
\eea
The spin correlation function can be evaluated in the usual way,
\be
C_9(r)= \langle s_z(r) s_z(0) \rangle  
\ee

\begin{figure}[t]
\begin{centering}
\includegraphics[width=0.45\columnwidth]{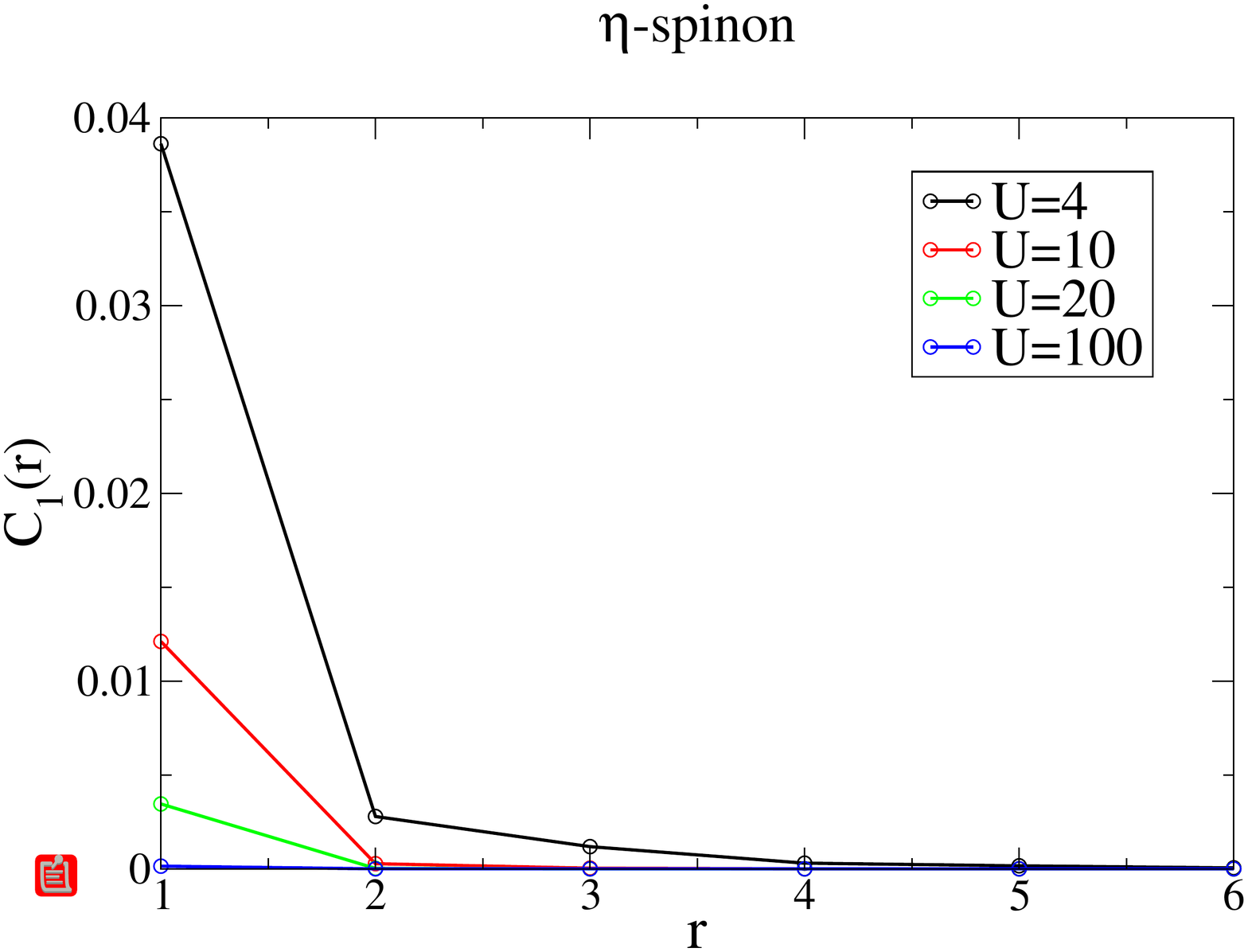}
\includegraphics[width=0.45\columnwidth]{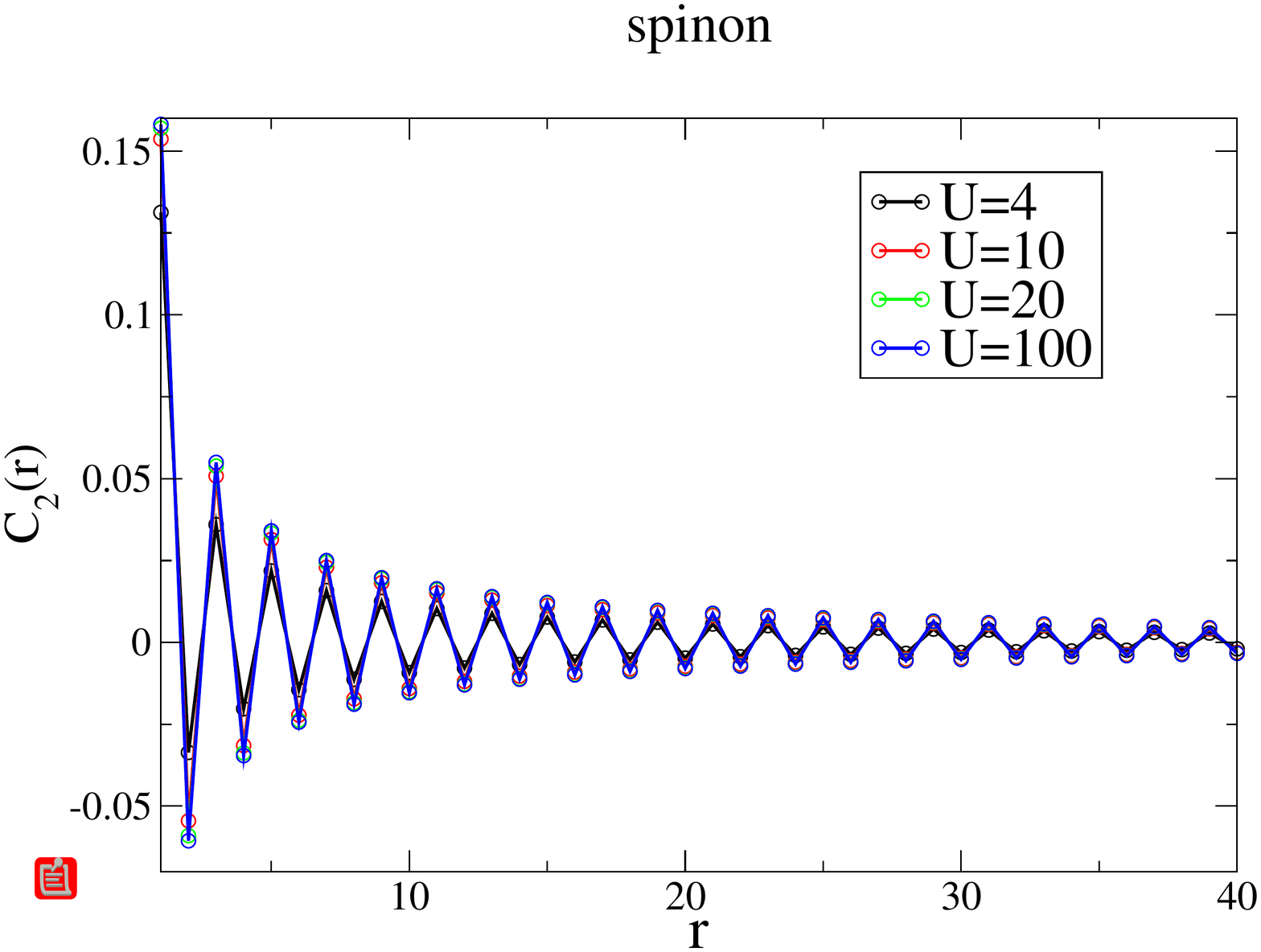}
\includegraphics[width=0.45\columnwidth]{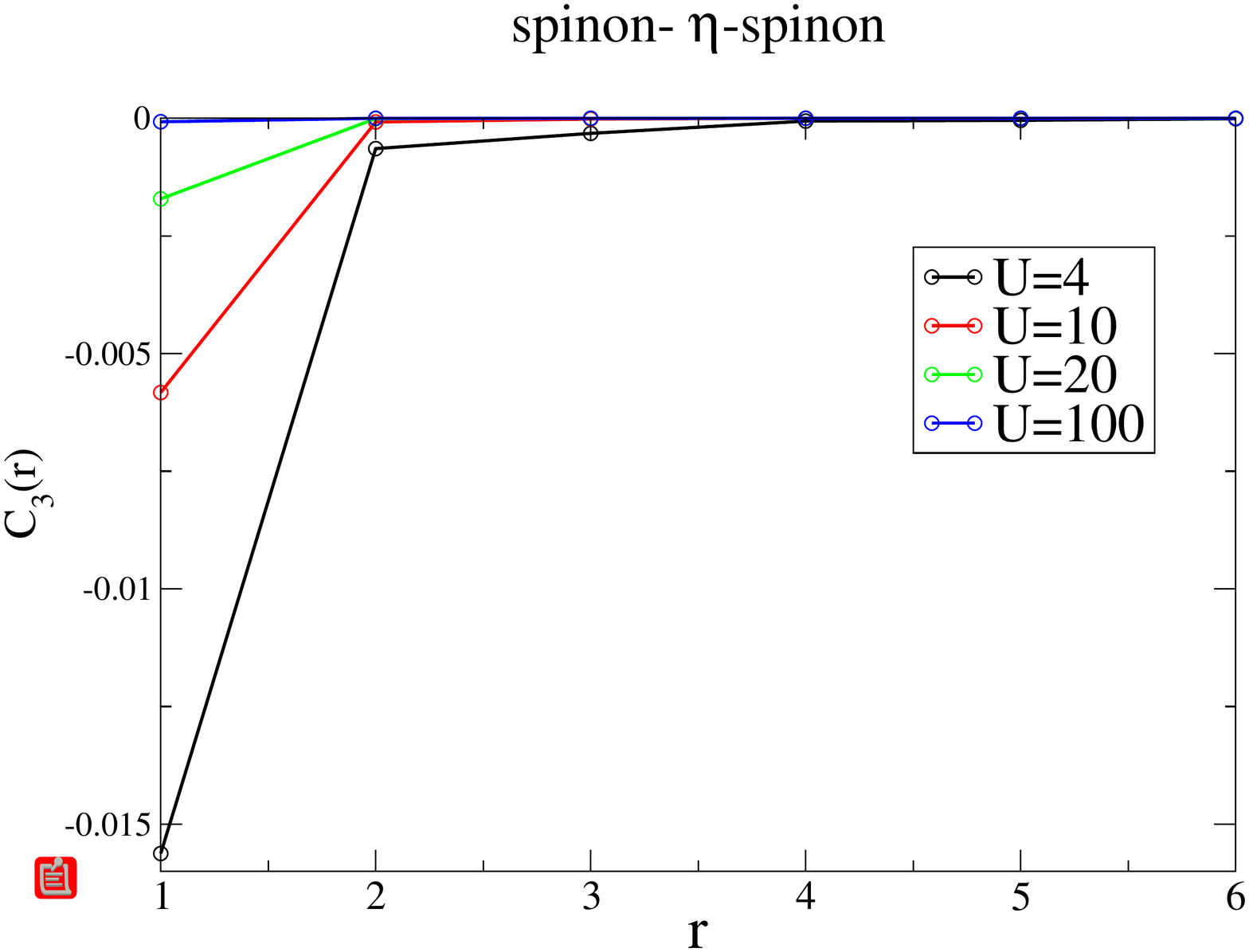}
\includegraphics[width=0.45\columnwidth]{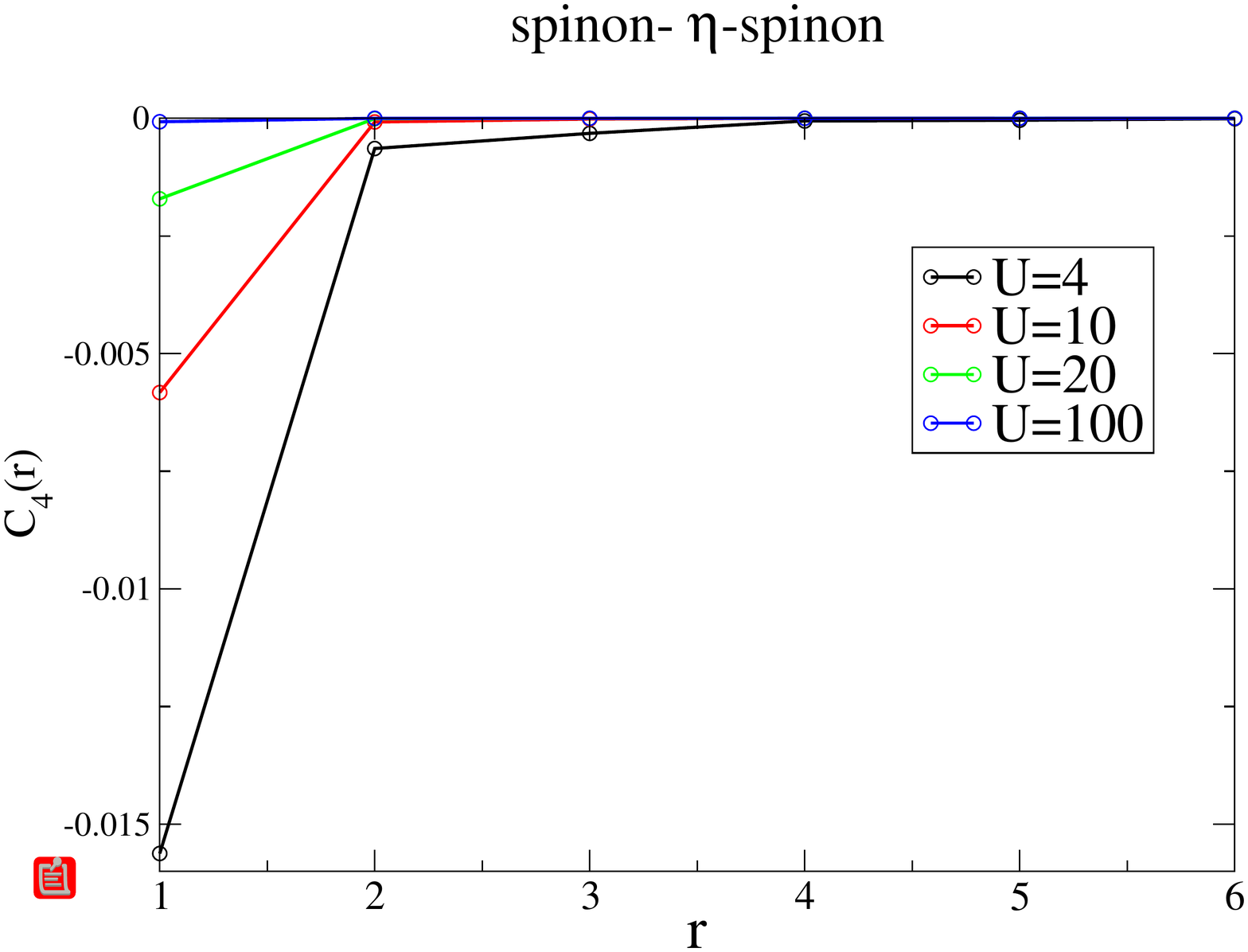}
\par\end{centering}
\caption{Connected correlation functions $C_1$ ($\eta$-spinon), $C_2$ (spinon),
$C_3$ and $C_4$ (spinon-$\eta$-spinon).
\label{figsc1-4} }
\end{figure}

A direct solution of these correlation functions in terms of the \"Ostlund and Granath
Hamiltonian is complicated, since even the mean-field approach is complex. Therefore,
we have used a DMRG method to calculate these and other correlation functions.

\subsection{DMRG calculations: Correlation functions}

For simplicity, here we limit ourselves to the half-filling case.
Using the DMRG method (briefly reviewed in Appendix B) we have calculated the correlation
functions indicated above. 
We consider correlation functions
for the original electron operators as a function of $U$. We emphasize that in the limit of very
large $U$ these equal the correlation functions of the rotated electrons for $U>0$. 

\begin{figure}[t]
\begin{centering}
\includegraphics[width=0.45\columnwidth]{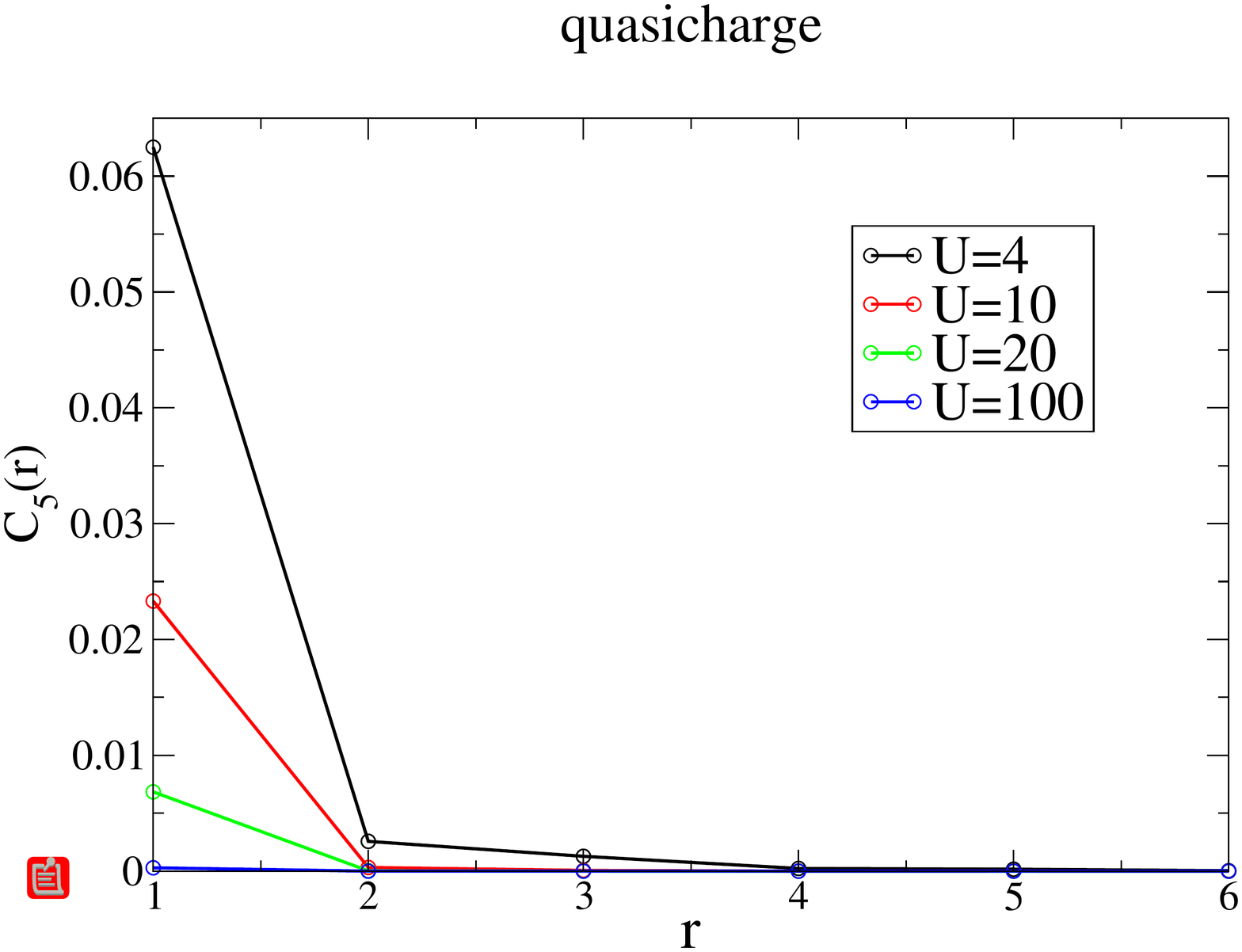}
\includegraphics[width=0.45\columnwidth]{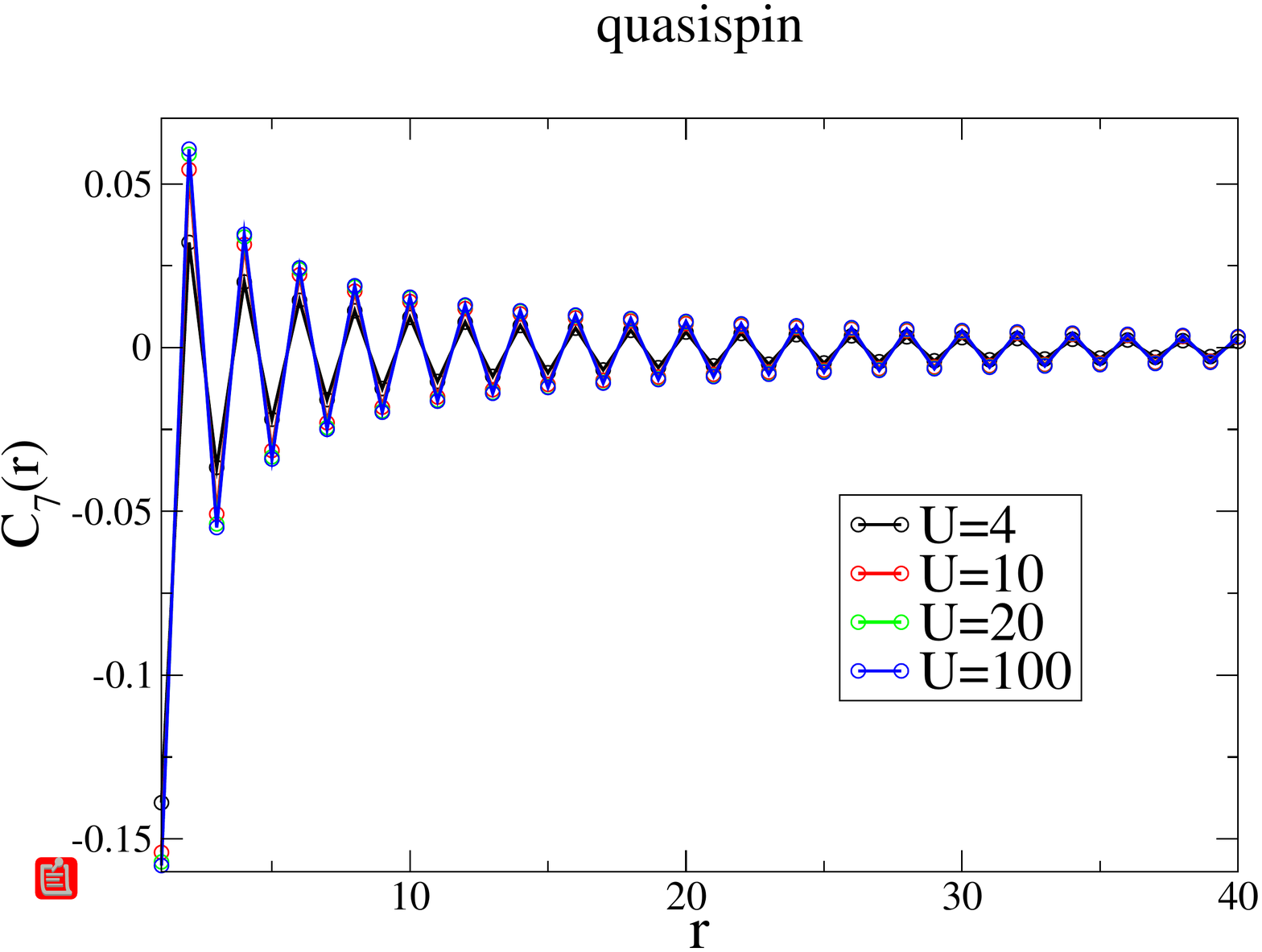}
\includegraphics[width=0.45\columnwidth]{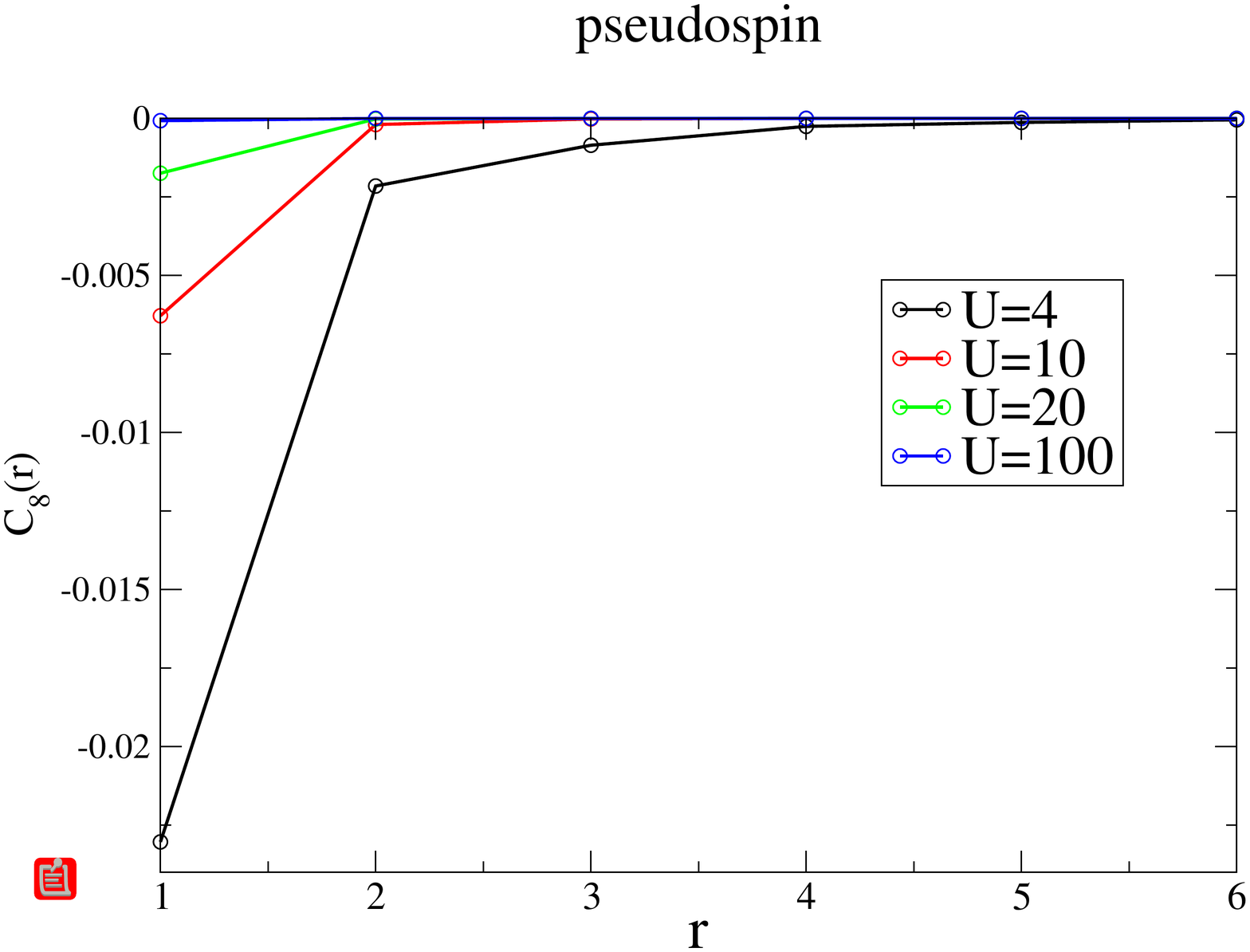}
\includegraphics[width=0.45\columnwidth]{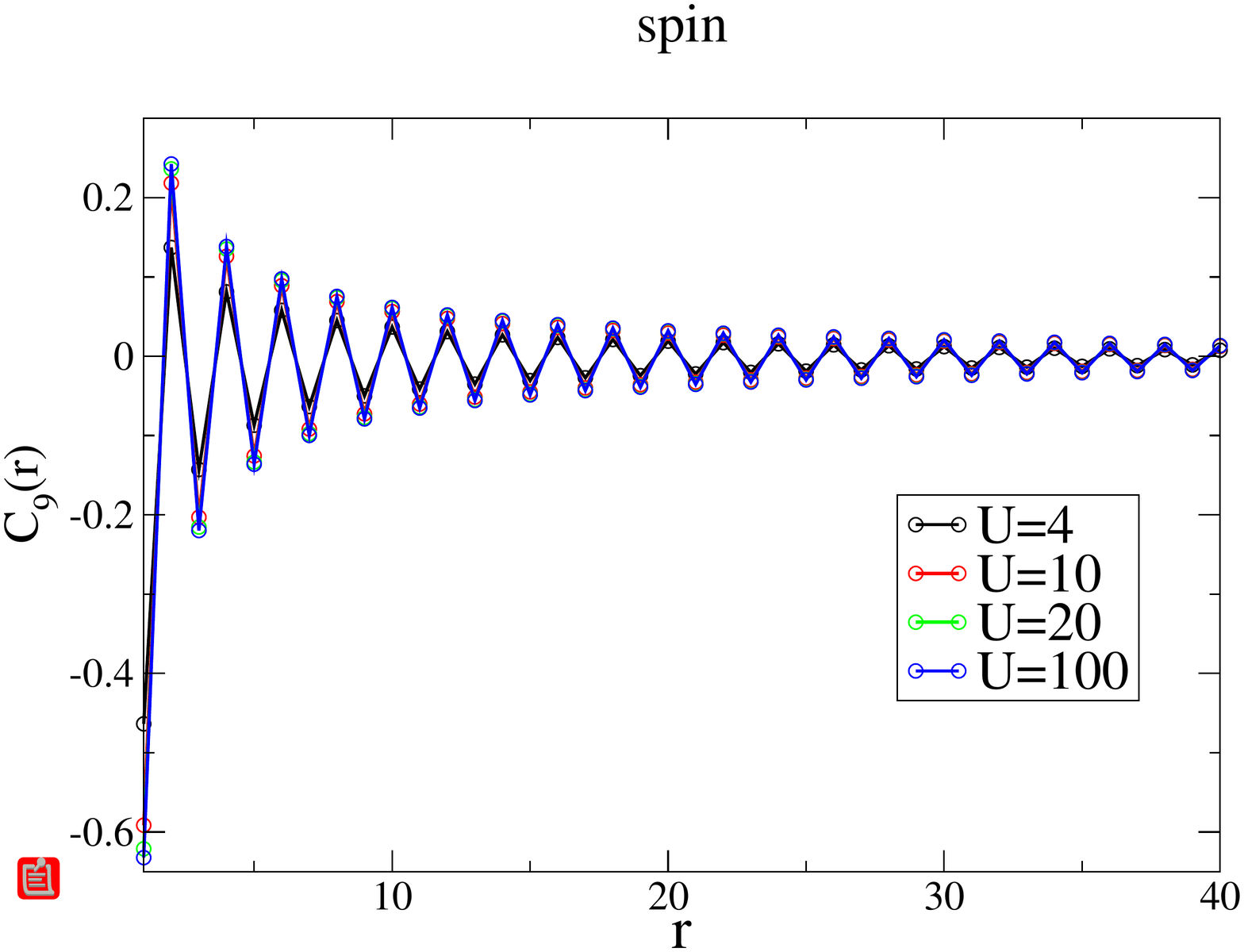}
\par\end{centering}
\caption{Connected correlation functions $C_5$ (quasicharge), $C_7$ (quasispin),
$C_8$ (pseudospin) and $C_9$ (spin).
\label{figsc5-9} }
\end{figure}

The charge correlation functions typically decay fast as the distance $r$ between
the two operators grows. On the other hand, the spin related correlation functions
oscillate by a factor of the type $(-1)^r$ and decay slowly with distance
with a power law behavior, like in the spin-$1/2$ isotropic Heisenberg model.

In Fig. (\ref{figsc1-4}) we present results for the correlations functions
$C_1$, $C_2$, $C_3$ and $C_4$. The correlation function $C_1(r)$ decays fast
with distance, which shows that the doubly occupied sites and the unoccupied sites
are tightly correlated in the half-filled phase. As $U$ grows the spatial extent
is strongly reduced and in the large $U$ limit the correlation is basically extended
to the nearest-neighbors. Since in the very large $U$ limit a doubly occupied site
costs an infinite energy, the correlation function basically vanishes.
The very large $U$ limit corresponds to the correlation function of the
rotated electrons for the whole $U>0$ range. Hence
this implies a very short range and a vanishing correlation function
in the strictly infinite limit.

The correlation functions for the doubly occupied sites and a singly occupied site
or a unoccupied site and a singly occupied site are negative. This is indicative of an
anti-correlation, as expected. These correlations also decay very fast with distance.

On the other hand, the correlation function $C_2(r)$ for a singly-occupied site
with spin up and a singly occupied site with opposite spin projection, is somewhat similar
to the longitudinal spin correlation function. It oscillates with distance and decays
slowly with $r$, which indicates a long-range correlation. The influence of the Hubbard interaction
is smaller than for the charge correlation functions. For large $U$ the short range values
of the correlation function increases, consistently with a more pronounced spin character
of the excitations of the half-filled Hubbard model at large interactions. However,
as $U$ becomes very large (for instance comparing $U=100$ with $U=20$) the decay with distance is faster. 

In Fig. (\ref{figsc5-9}) we show results for the correlation functions for the quasicharge,
quasispin and pseudospin operators introduced in Ref. \cite{Ostlund_2006}. The charge correlation
functions decay fast in a way similar to $C_1(r)$, specifically the quasicharge and the pseudospin
correlation functions. The quasicharge shows a correlation while the pseudospin shows an
anti-correlation. The two spin correlation functions have a slowly decaying oscillating
behavior. Analysis of both the quasispin and the spin correlation functions reveals that the nearest-neighbor
is anti-correlated. On the other hand, the spinon correlation function $C_2(r)$ behavior shows 
that the nearest-neighbors are positively correlated, since it is an occupation number correlation function. 
The other correlation functions respect the opposite spin projections.

The charge correlation functions can be fitted with an expression of the form given in Eq. (\ref{exponential}).
Both the decay length $\xi$ and the exponent $\sigma$ are indicative of a stronger decay as
the coupling grows. The results are shown in the Table.

\begin{equation*}
\begin{array}{lllll}
 & U=4 & U=10 & U=20 & U=100 \\
\sigma & 2.16 & 4.84 & 6.43 & 11.17 \\
\xi & 2.03 & 1.39 & 1. & 1. 
\end{array}
\label{table1}
\end{equation*}

The spin correlation functions can be fitted to an expression of the form provided in Eq. (\ref{power}).
The results for the various spin related correlation functions show that their decay is
very similar to the decay of the spin correlation function $C_9(r)$. In the infinite $U$ limit
these correlation functions tend to the corresponding correlation functions of the 
spin-$1/2$ isotropic Heisenberg model. 

\subsection{Eigenstates of the reduced density matrix}

The correlation functions we have calculated give information on the correlations between the various
entities discussed above. A more direct approach is obtained studying the eigenstates and eigenvalues
of the reduced density matrix of two sites in the chain. Considering a chain of $N$ sites we may
single out two sites distant by $r$ lattice units. The full density matrix of the chain can then be written as,
\be
\rho_N = |\psi \rangle \langle \psi |
\ee
where $|\psi \rangle $ is the ground state of the system. The ground state may be written as the direct
product of the states at each site. These can be written in terms of a basis with four states, namely
$|\phi \rangle = |0\rangle , |\uparrow \rangle , |\downarrow \rangle , |\uparrow \downarrow \rangle $, referring
to the four possibilities that each site is either unoccupied, occupied by a particle of spin up, a particle of
spin down or doubly occupied, respectively. The full density matrix is a $4^N \times 4^N$ matrix.
The ground state may be obtained for instance considering exact diagonalization of small
systems. We have used Lanczos method to obtain the ground state expressed in this basis. We considered
a system of size $N=14$.

Information about the correlation between two points on the lattice may be obtained considering
a reduced density matrix by integrating $N-2$ sites. One of the sites may be located at point $r=0$ on
the lattice and the other may be located at site $r$. The reduced density matrix is then
obtained as,
\be
\rho_2 = Tr_{N-2} |\psi \rangle \langle \psi |
\ee
This is a $16 \times 16$ matrix that can be diagonalized for different values of $U$ and different values
of $r$. The eigenvalues give the probabilities to find the two sites in a given correlated state
characterized by the corresponding eigenstate. 
We consider as before $U=4,10,20,100$ and $r=1,\cdots,7$.
In addition, we consider half-filling and zero magnetization, which implies seven electrons
with spin up and seven electrons with spin down in a chain with fourteen sites.

\vspace{\baselineskip}
\begin{equation*}
\begin{array}{ll}
\mbox{eigenstate} & |\phi_{r=0};\phi_r \rangle = |n_{0,\uparrow} n_{r,\uparrow} n_{0,\downarrow} n_{r,\downarrow} \rangle \\
A & 
\alpha \left(|1001 \rangle + |0110 \rangle \right)   +
\beta \left(|1010 \rangle + |0101 \rangle \right) \\
B & \frac{1}{\sqrt{2}} \left(|0110 \rangle - |1001 \rangle \right) \\
C & |1100 \rangle \\
D & |0011 \rangle \\
E & \frac{1}{\sqrt{2}} \left(|1011 \rangle + |0111 \rangle \right) \\
F & \frac{1}{\sqrt{2}} \left(|1110 \rangle + |1101 \rangle \right) \\
G & \frac{1}{\sqrt{2}} \left(|1000 \rangle + |0100 \rangle \right) \\
H & \frac{1}{\sqrt{2}} \left(|0010 \rangle + |0001 \rangle \right) \\
I & \frac{1}{\sqrt{2}} \left(|1011 \rangle - |0111 \rangle \right) \\
J & \frac{1}{\sqrt{2}} \left(|1110 \rangle - |1101 \rangle \right) \\
K & \frac{1}{\sqrt{2}} \left(|0010 \rangle - |0001 \rangle \right) \\
L & \frac{1}{\sqrt{2}} \left(|1000 \rangle - |0100 \rangle \right) \\
M & |1111 \rangle \\
N & |0000 \rangle \\
O & \frac{1}{\sqrt{2}} \left(|1010 \rangle - |0101 \rangle \right) \\
P & \gamma \left(|1001 \rangle + |0110 \rangle \right)   -
\delta \left(|1010 \rangle + |0101 \rangle \right)
\end{array}
\label{table2}
\end{equation*}

The structure of the normalized eigenstates is illustrated in the table. 
Those are linear combinations of a few states of the basis for sites $i,j$ represented as
$|n_{i,\uparrow} n_{j,\uparrow} n_{i,\downarrow} n_{j,\downarrow} \rangle$ where $n_i=0,1$ and $0$ means unoccupied 
and $1$ occupied. Since we are considering half filling, there is particle-hole symmetry.
Moreover, zero magnetization implies a symmetry between up and down spins.
The eigenstate with the highest eigenvalue is the state $A$ represented in the table.
This means it has the highest weight in the correlations between the two sites.

\begin{figure}[t]
\begin{centering}
\includegraphics[width=0.8\columnwidth]{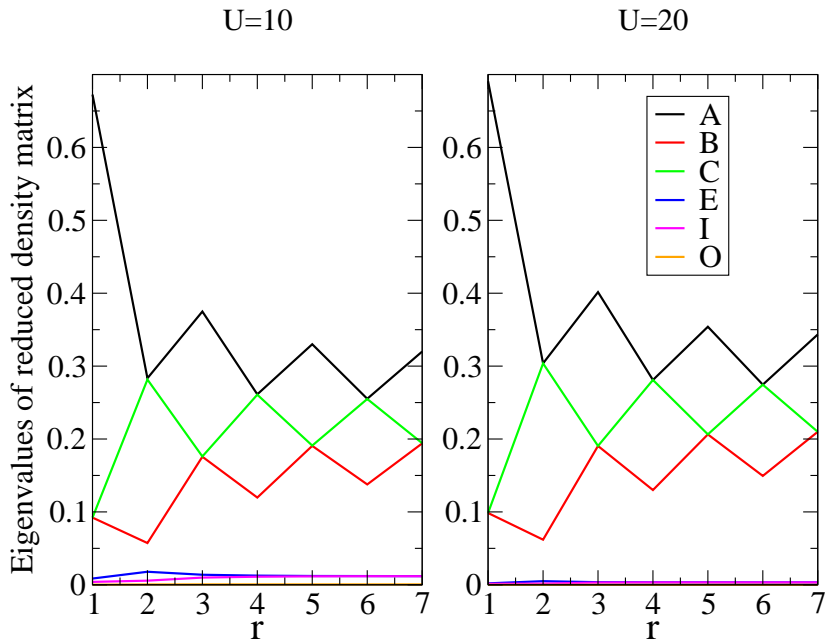}
\par\end{centering}
\caption{\label{eigens} 
Eigenvalues of the reduced density matrix, $\rho_2$, for $U=10$ (left panel) and $U=20$ (right
panel) as a function of distance $r$, for some eigenstates.
}
\end{figure}

State $A$ has four components with relative weights $\alpha,\beta$. The coefficient $\alpha$ measures the
contribution due to spins up and down in the two sites and the coefficient $\beta$ gives the contribution
of a double occupied site and a unoccupied site at $r=0$ and $r$. In Fig. \ref{eigens} we compare the weights
of some of the eigenstates as a function of the distance $r$ for $U=10,20$.
This state is spatially symmetric (exchanging the two sites) 
and anti-symmetric in spin space. It corresponds to a spin singlet.
The part of the state with weight $\alpha$ is a spin singlet between sites
$r=0$ and $r$ and the state with weight $\beta$ is a local spin singlet at either
site $r=0$ or site $r$ and a $\eta$-spin triplet by exchanging the doubly occupied site
with the unoccupied site. At half filling we expect
that for large U the state $A$ has a large weight, since it is associated with the formation
of spin singlets. 

State $O$ is associated with an antisymmetric pairing of a doubly occupied site with
a unoccupied site. It has a local spin singlet at the doubly occupied site.
This state has very small weight whereas state $P$ is the lowest-weight state.
It is the counterpart of state $A$.

\begin{figure}[t]
\begin{centering}
\includegraphics[width=0.8\columnwidth]{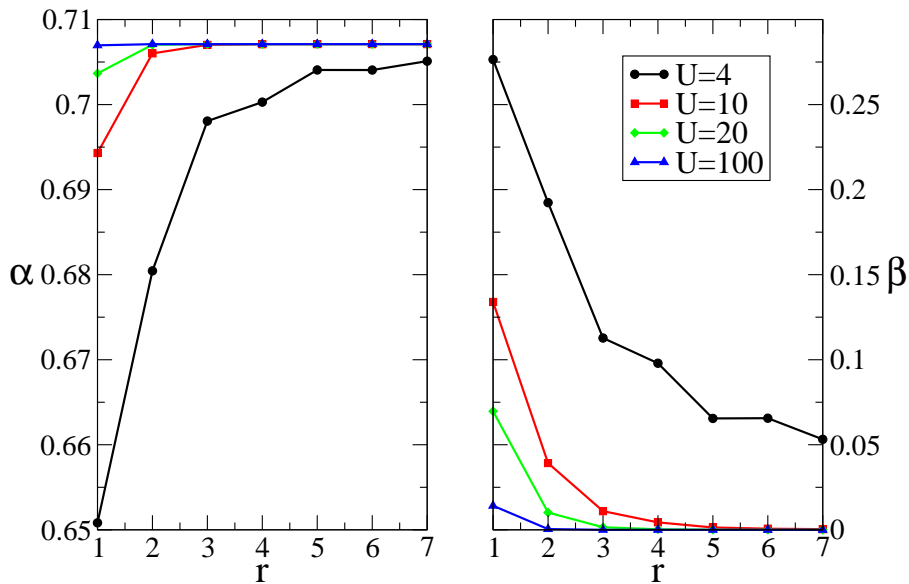}
\par\end{centering}
\caption{\label{amps} Relative weight of the spin up-spin - down-spin subspace $\alpha$ and doubly occupied-site - unoccupied-site 
subspace $\beta$ in the eigenstate with highest eigenvalue, $A$, for the matrix $\rho_2$. 
}
\end{figure}

States $B$ and $C$ (and its degenerate state $D$) also have a large weight. 
These states are spin triplets and antisymmetric in the space coordinates.
Interestingly states where the spins at points $r=0$ and $r$ are parallel have
equal or larger weight. For odd sites, states $B,C,D$ are degenerate but for even sites it is
more favorable for the spins to be parallel. This is indicative of the long-range antiferromagnetic
correlations in the large $U$ limit at half-filling. 

States $E$ and $I$ are representative of mixtures between states involving singly occupied
sites and doubly or unoccupied sites, either in antisymmetric or symmetric combinations, and have smaller weights. 

Entanglement between doubly occupied sites and unoccupied sites enters indirectly via
state $A$, since state $O$ has very low weight. The lowest probability state $P$ also has a mixture of the same four states,
as well with a very small weight.

Even though the doubly occupied-site - unoccupied-site entanglement contributes to state $A$, the relative
weight decreases fast with distance and interaction strength. This is illustrated in Fig. \ref{amps}.
In the left panel we show the relative weight of the spin up-spin - down-spin subspace $\alpha$ and in the
right panel we show that of the doubly-occupied-site - unoccupied-site subspace $\beta$. Note that 
the dominant contribution comes from the spin up-spin - down-spin subspace $\alpha>\beta$. (As a side
remark, the eigenstate with the lowest eigenvalue, $P$, has $\gamma < \delta$.)
Comparing the various values of the interaction strength $U$, we see that $\beta$ decreases fast.
The same happens as $r$ increases, which is consistent with the short range of the correlations
between a doubly occupied site and a unoccupied site. For large $U$ the weight is mostly contained
in the spin part, in a spin-singlet state. These results are consistent and clarify the previous
results that the charge correlations are very short range, particularly as $U$ grows.

Tracing out all states except those with unoccupied or doubly occupied sites, we find a $4\times 4$
matrix that stores direct information on the correlations between a unoccupied site and a doubly occupied site.
This density matrix may be defined as,
\be
\rho_3= Tr^{\prime} \rho_2
\ee
where the trace is over the singly occupied sites.

\begin{figure}[t]
\begin{centering}
\includegraphics[width=0.8\columnwidth]{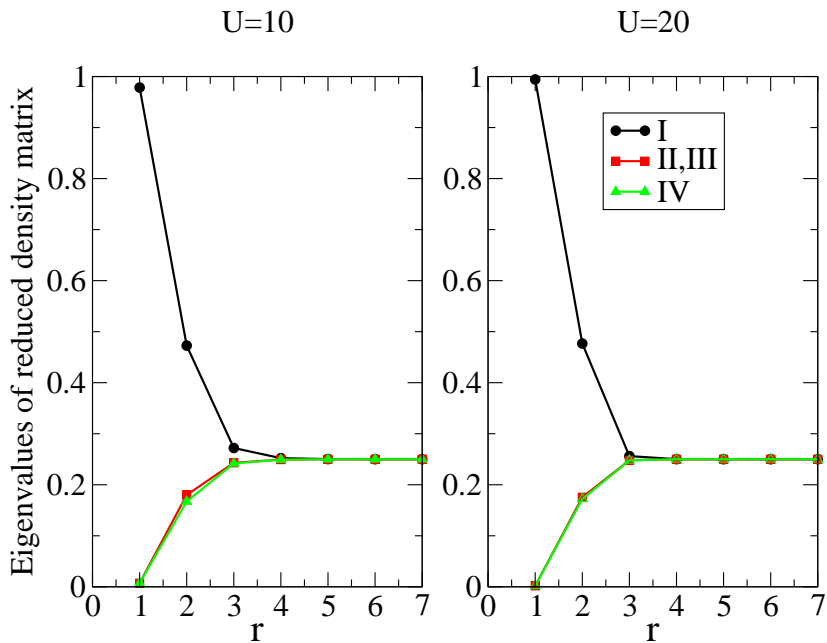}
\par\end{centering}
\caption{\label{subspace} 
Eigenvalues of the reduced density matrix $\rho_3$ for $U=10$ (left panel) and $U=20$ (right
panel), as a function of distance $r$.
}
\end{figure}

The eigenstates of this reduced density matrix are of the form,

\vspace{\baselineskip}
\begin{equation*}
\begin{array}{ll}
\mbox{eigenstate} & |\phi_{r=0};\phi_r \rangle = |n_{0,\uparrow} n_{r,\uparrow} n_{0,\downarrow} n_{r,\downarrow} \rangle \\
I & \frac{1}{\sqrt{2}} \left(|1010 \rangle + |0101 \rangle \right) \\
II & |1111 \rangle \\
III & |0000 \rangle \\
IV & \frac{1}{\sqrt{2}} \left(|1010 \rangle - |0101 \rangle \right)
\end{array}
\label{table2}
\end{equation*}
The results for the eigenvalues for $U=10,20$ of this reduced density matrix are shown in Fig. \ref{subspace}.
The eigenvalue coresponding to state $I$ that mixes a doubly occupied site with a unoccupied site is the largest
at small distances and as distance increases all states become equally probable. The decrease of the relative
weight of state $I$ increases with $U$, consistent with previous results. This state is a $\eta$-spin triplet
(note that this is consistent with the structure in Eq. \ref{mftH}).
The lowest-weight state is a $\eta$-spin singlet. The degenerate states $II$ and $III$ correspond to two
doubly occupied sites and two unoccupied sites.

For the subspace spanned by $\{|0>, |\uparrow, \downarrow>\}\times \{|0>, |\uparrow, \downarrow>\}$  
defined on two sites, the entanglement can be measured by the concept
of concurrence \cite{wootters}. 
While the concurence is still a combination of some correlation functions, it is different from the traditional
density-density or other types of correlation function. The entanglement results from the linear superposition principle of the
quantum mechanics and is absent in the classical physics. Therefore, it is usually regarded as a kind of pure quantum correlation.

The reduced density matrix can be written as,
\begin{equation}
\rho_3 = \left(
\begin{array}{llll}
u & 0 & 0 & 0 \\
0 & w_1 & z & 0 \\
0 & z^* & w_2 & 0 \\
0 & 0 & 0 &v
\end{array} \right)
\end{equation}
The concurrence, as the measure of the entanglement, can be calculated as,
\be
C = 2 \mbox{max}[0,|z|-\sqrt{u v} ]
\ee
The results for the concurrence show that both for $U=10$ and
$U=20$ the correlations only extend to nearest-neighbors. 
The concurrence for higher values of the distance between the two sites $r>1$ vanishes. For nearest
neighbors the concurrence takes the values $C=0.9577$ and $C=0.9884$ for 
$U=10$ and $U=20$, respectively, showing the large increase for large $U$ and that the correlation is strong,
since the concurrence is close to $1$.

\section{Summary}

In this paper we have studied the correlation functions of basic
entities of the one-dimensional Hubbard model using various methods.
Previous analysis of the exact solution via the Bethe ansatz 
suggests the importance of correlations between doubly occupied and unoccupied
sites and sites singly occupied with spin up and spin down electrons. 
These correlations have also been suggested by other treatments, as mentioned
in the text. 

The relevance of these operators is also stressed by their connection in terms
of so-called rotated electrons with basic entities of the exact solution such as
spinons and $\eta$-spinons. We have, therefore, calculated various correlation
functions using an approximate mean-field solution within the Zou-Anderson
transformation and with the introduction of non-local bond variables, in a way
that is a reminder of the spinon and $\eta$-spinon bound and anti-bound states,
respectively. Furthermore, we have used
exact DMRG calculations of the same quantities and supplemented those with calculations
of correlation functions for the operators introduced by \"Ostlund and Granath,
and further developed by one of the authors, which allow an exact charge-spin
separation in a way that reminds the exact separation in low-dimensional
systems.

We concluded that the charge-like correlation functions are typically very short ranged
in the case of half filling, due to the charge gap. As the interaction strength
increases, the correlations become virtually nearest-neighbor like. The spin-like
correlation functions are however more extended in a way similar to the spin-spin
correlation functions of the Hubbard model or, in the large $U$ limit, those of
the spin-$1/2$ isotropic Heisenberg model. Even though the correlation functions calculated here are
different from a standard longitudinal spin correlation function, their decay with
distance is similar.

Further insight onto the correlations between two sites in a chain was obtained
calculating the eigenvalues and eigenstates of the two-site reduced density
matrix using exact diagonalization of a small system, also at half filling.
The role of the spin and charge contributions to the entanglement between the
two sites was clarified. As shown by the other methods, the correlations between
doubly occupied sites and unoccupied sites are very short ranged, as evidenced by the
concurrence which extends only to nearest neighbors. The eigenstates of the reduced density matrix
corresponding to singly occupied sites with spin up or spin down have longer range.

The results using the mean-field approach were also extended to cases in the metallic
phase (away from half-filling). The absence of the charge gap leads to
correlations between a doubly occupied site and a unoccupied site that have larger
range. The mean-field treatment also allows the study of higher energy phases. It
was shown that for some of these phases the charge correlations have comparable
ranges to those of the spin correlations.

\ack
We thank Pedro Ribeiro, Miguel Ara\'ujo, Peter Horsch and Alejandro Muramatsu for discussions and the hospitality and support of the
Beijing Computational Science Research Center. J.M.P.C. thanks the hospitality of the Institut f\"ur Theoretische Physik III, Universit\"at Stuttgart and 
support by the Portuguese FCT under SFRH/BSAB/1177/2011, German transregional collaborative research center SFB/TRR21, and
Max Planck Institute for Solid State Research.

\appendix

\section{Mean-field solution using the Zou, Anderson transformation}

Here we briefly review the mean-field solution of the Hubbard model in terms
of the link variables introduced in Eq. (\ref{mftH}).
The bosonic and fermionic parts decouple. The mean-field Hamiltonian may be written as,
\be
H_{MF} = H_{ed} + H_{SS} + C
\ee
where
\bea
H_{ed} = &-& t \sum_{i,\delta} \left\{ \left( e_{i+\delta}^{\dagger} e_i -d_{i+\delta}^{\dagger}
d_i \right) \left( \chi_{\delta,\uparrow}^S + \chi_{\delta,\downarrow}^S \right)
+ e_i d_{i+\delta} \Delta_{\delta}^* +d_{i}^{\dagger} e_{i+\delta}^{\dagger} \Delta_{\delta} \right\}
\nonumber \\
&+& \sum_i d_i^{\dagger} d_i \left( U -\mu +\lambda_i \right)
+ \sum_i e_i^{\dagger} e_i \left( \mu + \lambda_i \right)
\eea
\bea
H_{SS} = &-& t \sum_{i,\delta} \left\{ \left( \chi_{\delta}^e - \chi_{\delta}^d \right) \sum_{\sigma}
S_{i,\sigma}^{\dagger} S_{i+\delta,\sigma} 
+ \Phi_{\delta} \left( S_{i,\uparrow}^{\dagger}
S_{i+\delta,\downarrow}^{\dagger} - S_{i,\downarrow}^{\dagger} S_{i+\delta,\uparrow}^{\dagger} \right) \right.
\nonumber \\
&+& \left. \Phi_{\delta}^* \left( S_{i,\downarrow}
S_{i+\delta,\uparrow} - S_{i,\uparrow} S_{i+\delta,\downarrow} \right) \right\}
\nonumber \\
&+& \sum_{i,\sigma} \lambda_i  S_{i,\sigma}^{\dagger} S_{i,\sigma}
\eea
\bea
C= &t& \sum_{i,\delta,\sigma} \left\{ \left( \chi_{\delta}^e - \chi_{\delta}^d \right) \chi_{\delta,\sigma}^S
+ \Phi_{\delta} \Delta_{\delta}^* + \Phi_{\delta}^* \Delta_{\delta} \right\} \nonumber \\
&-& \mu N - \sum_i \lambda_i
\eea

Defining the Fourier transforms of the operators as,
\be
e_r = \frac{1}{\sqrt{N_a}} \sum_k e^{i k r} e_k
\ee
and
\be
\chi_{k}^A = \sum_{\delta} e^{-i k \delta} \chi_{\delta}^A
\ee
where $A=e,d,S$, and similarly for $\Phi_k$ and $\Delta_k$, 
we can write in momentum space that,
\bea
H_{ed} &=& \sum_k \left\{ \left( -t \sum_{\sigma} \chi_{k,\sigma}^S + \mu + \lambda_0 \right)
e_k^{\dagger} e_k + \left( t \sum_{\sigma} \chi_{k,\sigma}^S +U-\mu +\lambda_0 \right)
d_k^{\dagger} d_k \right. \nonumber \\
&+& \left. \left( -t \Delta_k^* e_k d_{-k} - t \Delta_k d_k^{\dagger} e_{-k}^{\dagger} \right) \right\}
\eea
That is,
\be
H_{ed} = \sum_k \left\{ E_k e_k^{\dagger} e_k +D_k d_k^{\dagger} d_k 
-t \Delta_k^* e_k d_{-k}
- t \Delta_k d_k^{\dagger} e_{-k}^{\dagger}  \right\}
\ee
where
\bea
E_k &=& -t \sum_{\sigma} \chi_{k,\sigma}^S + \mu + \lambda_0 \nonumber \\
D_k &=& t \sum_{\sigma} \chi_{k,\sigma}^S +U-\mu +\lambda_0 \nonumber 
\eea
and
\bea
H_{SS} &=& \sum_k \left\{ \sum_{\sigma} \left( -t (\chi_k^e - \chi_k^d) + \lambda_0 \right)
S_{k,\sigma}^{\dagger} S_{k,\sigma} \right. \nonumber \\
&-& t \left. \Phi_k \left( 
S_{k,\uparrow}^{\dagger} S_{-k,\downarrow}^{\dagger} -
S_{k,\downarrow}^{\dagger} S_{-k,\uparrow}^{\dagger} \right)
-t \Phi_k^* \left( 
S_{k,\downarrow} S_{-k,\uparrow} -
S_{k,\uparrow} S_{-k,\downarrow} \right) \right\}
\eea
One then arrives to,
\bea
H_{SS} &=& \sum_{k,\sigma} \left\{ \bar{\epsilon}_k S_{k,\sigma}^{\dagger} S_{k,\sigma}
- t \Phi_k \left(
S_{k,\uparrow}^{\dagger} S_{-k,\downarrow}^{\dagger} -
S_{k,\downarrow}^{\dagger} S_{-k,\uparrow}^{\dagger} \right) \right. \nonumber \\
&-& \left. t \Phi_k^* \left(
S_{k,\downarrow} S_{-k,\uparrow} -
S_{k,\uparrow} S_{-k,\downarrow} \right) \right\}
\eea
where
\be
\bar{\epsilon}_k = -t (\chi_k^e - \chi_k^d) + \lambda_0 \nonumber
\ee
We have assumed that the Lagrange multiplier is uniform and only the $k=0$ component
is non-vanishing. For later purposes we define,
\[
\gamma_k = \sum_{\delta} e^{-i k \cdot \delta}
\]

We are left with the diagonalization of two quadratic Hamiltonians. This can be done in a
standard way performing Bogoliubov-Valatin transformations. 
Defining,
\bea
e_k &=& u_k \alpha_k + v_k^* \beta_{-k}^{\dagger} \nonumber \\
d_{-k} &=& u_k \beta_{-k} + v_k^* \alpha_k^{\dagger}
\eea
and similarly for the spin part,
\bea
s_{\uparrow,k} &=& \tilde{u}_k \tilde{\alpha}_k + \tilde{v}_k^* \tilde{\beta}_{-k}^{\dagger} \nonumber \\
s_{\downarrow,-k} &=& \tilde{u}_k \tilde{\beta}_{-k} - \tilde{v}_k^* \tilde{\alpha}_{k}^{\dagger} 
\eea
where $u_k^2-v_k^2=1$ for bosons and $\tilde{u}_k^2+\tilde{v}_k^2=1$ for fermions,
and eliminating off-diagonal terms in the quasiparticle operators, we obtain that
the diagonalized Hamiltonians take the forms,
\bea
H_{ed} &=& \sum_k \left\{ \epsilon_k^{\alpha} \alpha_k^{\dagger} \alpha_k + \epsilon_k^{\beta}
\beta_{-k}^{\dagger} \beta_{-k} + [\epsilon_k-1/2 (E_k+D_k)] \right\} \nonumber \\
H_{SS} &=& \sum_k \left\{ \tilde{\epsilon}_k \left( \tilde{\alpha}_k^{\dagger} 
\tilde{\alpha}_k + 
\tilde{\beta}_{-k}^{\dagger} \tilde{\beta}_{-k} \right)+
[\bar{\epsilon}_k-\tilde{\epsilon}_k] \right\}
\eea
where
\bea
\epsilon_k^{\alpha} &=&  \epsilon_k + 1/2 (E_k-D_k)
\nonumber \\
\epsilon_k^{\beta} &=& \epsilon_k - 1/2 (E_k-D_k)
\eea
with
\be
\epsilon_k = \sqrt{\left( \frac{E_k+D_k}{2} \right)^2 - t^2 \Delta_k^2}
\ee
Note that $E_k+D_k=U+2 \lambda_0$.
Here,
\be
\tilde{\epsilon}_k = \sqrt{ \bar{\epsilon}_k^2 + 4 t^2 \Phi_k^2 }
\ee
Also, we find that,
\bea
u_k v_k &=& \frac{t \Delta_k}{2 \epsilon_k} \nonumber \\
u_k^2 &=& \frac{1}{2} \left\{ \frac{E_k+D_k}{2 \epsilon_k} +1 \right\} \nonumber \\
v_k^2 &=& \frac{1}{2} \left\{ \frac{E_k+D_k}{2 \epsilon_k} -1 \right\}
\eea
and
\bea
\tilde{u}_k \tilde{v}_k &=& \frac{t \Phi_k}{\tilde{\epsilon}_k} \nonumber \\
\tilde{u}_k^2 &=& \frac{1}{2} \left\{ 1+ \frac{\bar{\epsilon}_k}{\tilde{\epsilon}_k} 
\right\} \nonumber \\
\tilde{v}_k^2 &=& \frac{1}{2} \left\{ 1- \frac{\bar{\epsilon}_k}{\tilde{\epsilon}_k}
\right\} 
\eea
 
The solution of the problem involves the self-consistent calculation of the averages appearing
in these equations. Any average can now be calculated using the Bogoliubov-Valatin transformations.

The mean-field equations are given by,
\be
\Delta_k = \frac{2}{N_s} \sum_{k'} \gamma_{k-k'}
\left\{ \tilde{u}_{k'} \tilde{v}_{k'} \left( 1-2 f(\tilde{\epsilon}_{k'}) \right)
\right\}
\ee
\bea
\Phi_k = \frac{1}{N_s} \sum_{k'} \gamma_{k-k'}
\left\{ \right.
&+&\left. u_{k'} v_{k'} \left( 1+\delta_{k',k_{\alpha}} N_s n_{\alpha} +
(1-\delta_{k',k_{\alpha}}) f_B(\epsilon_{k'}^{\alpha}) \right) \right. \nonumber \\
&+& \left. u_{k'} v_{k'} \left( \delta_{k',k_{\beta}} N_s n_{\beta} +
(1-\delta_{k',k_{\beta}}) f_B(\epsilon_{k'}^{\beta}) \right)
\right\}
\eea
\be
\chi_{k,\sigma}^S = \frac{1}{N_s} \sum_{k'} \gamma_{k-k'}
\left\{ \tilde{u}_{k'}^2 f(\tilde{\epsilon}_{k'})
+ \tilde{v}_{k'}^2 \left(1-f(\tilde{\epsilon}_{k'}) \right) 
\right\}
\ee
\bea
\chi_k^e = \frac{1}{N_s} \sum_{k'} \gamma_{k-k'}
\left\{ \right.
&+&\left. u_{k'}^2 \left( \delta_{k',k_{\alpha}} N_s n_{\alpha} +
(1-\delta_{k',k_{\alpha}}) f_B(\epsilon_{k'}^{\alpha}) \right) \right. \nonumber \\
&+& \left. v_{k'}^2 \left(1+ \delta_{k',k_{\beta}} N_s n_{\beta} +
(1-\delta_{k',k_{\beta}}) f_B(\epsilon_{k'}^{\beta}) \right)
\right\}
\eea
\bea
\chi_k^d = \frac{1}{N_s} \sum_{k'} \gamma_{k-k'}
\left\{ \right.
&+&\left. u_{k'}^2 \left( \delta_{k',k_{\beta}} N_s n_{\beta} +
(1-\delta_{k',k_{\beta}}) f_B(\epsilon_{k'}^{\beta}) \right) \right. \nonumber \\
&+& \left. v_{k'}^2 \left(1+ \delta_{k',k_{\alpha}} N_s n_{\alpha} +
(1-\delta_{k',k_{\alpha}}) f_B(\epsilon_{k'}^{\alpha}) \right)
\right\}
\eea
\bea
1 = \frac{1}{N_s} \sum_{k} 
\left\{ \right.
&+&\left. u_{k}^2 \left( \delta_{k,k_{\alpha}} N_s n_{\alpha} +
(1-\delta_{k,k_{\alpha}}) f_B(\epsilon_{k}^{\alpha}) \right) \right. \nonumber \\
&+& \left. v_{k}^2 \left(1+ \delta_{k,k_{\beta}} N_s n_{\beta} +
(1-\delta_{k,k_{\beta}}) f_B(\epsilon_{k}^{\beta}) \right)
\right. \nonumber \\
&+&\left. u_{k}^2 \left( \delta_{k,k_{\beta}} N_s n_{\beta} +
(1-\delta_{k,k_{\beta}}) f_B(\epsilon_{k}^{\beta}) \right) \right. \nonumber \\
&+& \left. v_{k}^2 \left(1+ \delta_{k,k_{\alpha}} N_s n_{\alpha} +
(1-\delta_{k,k_{\alpha}}) f_B(\epsilon_{k}^{\alpha}) \right)
\right. \nonumber \\
&+& \left. 2 \tilde{u}_k^2 f(\tilde{\epsilon}_k) + 2 \tilde{v}_k^2
\left( 1- f(\tilde{\epsilon}_k) \right) \right\}
\eea
\be
n-1 = -\left( n_{\alpha} + \sum_{k \ne k_{\alpha}} f_B (\epsilon_{k}^{\alpha}) \right)
+ \left( n_{\beta} + \sum_{k \ne k_{\beta}} f_B (\epsilon_{k}^{\beta}) \right)
\ee
where $f$ is the Fermi-Dirac distribution and $f_B$ the Bose-Einstein distribution.

The correlation functions may be calculated at the mean-field level.
We obtain that,
\bea
& & C_1(r) = \frac{1}{N_a^2} \sum_{k_1} \sum_{k_2}  \nonumber \\
& & \left[ |u_{k_1}|^2 |u_{k_2}|^2  
+ e^{-i(k_1-k_2)r} u_{k_1}^* v_{k_1} u_{k_2} v_{k_2}^* \right] f_B(\epsilon_{k_1}^{\alpha}) f_B(\epsilon_{k_2}^{\beta}) \nonumber \\
&+& \left[ |u_{k_1}|^2 |v_{k_2}|^2  
+ e^{-i(k_1-k_2)r} u_{k_1}^* v_{k_1} u_{k_2} v_{k_2}^* \right] f_B(\epsilon_{k_1}^{\alpha}) (1+f_B(\epsilon_{k_2}^{\alpha}))  \nonumber \\
&+& \left[ |v_{k_1}|^2 |u_{k_2}|^2 
+ e^{-i(k_1-k_2)r} u_{k_1}^* v_{k_1} u_{k_2} v_{k_2}^* \right] f_B(\epsilon_{k_2}^{\beta}) (1+f_B(\epsilon_{k_1}^{\beta})) \nonumber \\
&+& \left[ |v_{k_1}|^2 |v_{k_2}|^2 
+  e^{-i(k_1-k_2)r} u_{k_1}^* v_{k_1} u_{k_2} v_{k_2}^* \right] (1+f_B(\epsilon_{k_1}^{\beta})) (1+f_B(\epsilon_{k_2}^{\alpha})) 
\eea
and
\bea
\langle n_e(r) \rangle &=& \frac{1}{N_a} \sum_k \left[ |u_k|^2 f_B(\epsilon_k^{\alpha}) + |v_k|^2 (1+f_B(\epsilon_k^{\beta})) \right]
\nonumber \\
\langle n_d(r) \rangle &=& \frac{1}{N_a} \sum_k \left[ |u_k|^2 f_B(\epsilon_k^{\beta}) + |v_k|^2 (1+f_B(\epsilon_k^{\alpha})) \right]
\eea
Similarly,
\bea
& & C_2(r) = \frac{1}{N_a^2} \sum_{k_1} \sum_{k_2}  \nonumber \\
& & \left[|\tilde{u}_{k_1}|^2 |\tilde{u}_{k_2}|^2  
+ e^{i(k_1-k_2)r} \tilde{u}_{k_1}^* \tilde{v}_{k_1} \tilde{u}_{k_2} \tilde{v}_{k_2}^* \right] 
f(\epsilon_{k_1}^{\beta}) f(\epsilon_{k_2}^{\alpha}) \nonumber \\
&+& \left[ |\tilde{u}_{k_1}|^2 |\tilde{v}_{k_2}|^2  
- e^{i(k_1-k_2)r} \tilde{u}_{k_1}^* \tilde{v}_{k_1} \tilde{u}_{k_2} \tilde{v}_{k_2}^* \right] 
f(\epsilon_{k_1}^{\beta}) (1-f(\epsilon_{k_2}^{\alpha}))  \nonumber \\
&+& \left[ |\tilde{v}_{k_1}|^2 |\tilde{u}_{k_2}|^2 
- e^{i(k_1-k_2)r} \tilde{u}_{k_1}^* \tilde{v}_{k_1} \tilde{u}_{k_2} \tilde{v}_{k_2}^* \right] 
f(\epsilon_{k_2}^{\alpha}) (1-f(\epsilon_{k_1}^{\alpha})) \nonumber \\
&+& \left[ |\tilde{v}_{k_1}|^2 |\tilde{v}_{k_2}|^2 
+  e^{i(k_1-k_2)r} \tilde{u}_{k_1}^* \tilde{v}_{k_1} \tilde{u}_{k_2} \tilde{v}_{k_2}^* \right] 
(1-f(\epsilon_{k_1}^{\alpha})) (1-f(\epsilon_{k_2}^{\beta})) 
\eea
and
\bea
\langle n_{s,\uparrow}(r) \rangle &=& \frac{1}{N_a} \sum_k \left[ |\tilde{u}_k|^2 f(\epsilon_k^{\beta}) + 
|\tilde{v}_k|^2 (1-f(\epsilon_k^{\alpha})) \right]
\nonumber \\
\langle n_{s,\downarrow}(r) \rangle &=& \frac{1}{N_a} \sum_k \left[ |\tilde{u}_k|^2 f(\epsilon_k^{\alpha}) + 
|\tilde{v}_k|^2 (1-f(\epsilon_k^{\beta})) \right]
\eea

\section{DMRG method}

In this Appendix a brief description of the DMRG method used in our
studies is presented. The density matrix renormalization group (DMRG)~\cite{SRW99}
algorithm is an accurate method in dealing with
quasi-one-dimensional system. It provides a criterion to find which states to
keep and which to discard. Therefore, it can deal with relative
large-size system with high accuracy. In the following, we give a
brief introduction about this method and its application to our
calculations.

In this method one constructs a superblock composed of the original system block and
the environment block,
usually the reflection of the system block.
The reduced density matrix for the system
block is defined as,
\be
\rho_{ii^\prime}=\sum_j{\psi_{ij}^* \psi_{i^\prime j} }.\label{eq:1}
\ee
where $\psi$ is a state of the superblock. Usually it is chosen as
the ground state corresponding to the zero temperature. Moreover, here $\left|
i\right\rangle$ and $\left| j\right\rangle$ label the states of the
system and the environment blocks, respectively. For any system
block operator $A$, we have,
\be
\left\langle
A\right\rangle=\sum_{ii^\prime}A_{ii^\prime}\rho_{ii^\prime}=Tr\left(
\rho A\right)=\Sigma_\alpha \omega_\alpha\left\langle
u^\alpha\left|A\right|u^\alpha\right\rangle.\label{eq:1}
\ee
where ${\omega_\alpha}$ and $u^\alpha$ are the eigenvalues and
eigenstates of the reduced matrix $\rho$. The
significance of the state $u^\alpha$ can then be determined by
$\omega_\alpha$. For a certain $\alpha$, if $\omega_\alpha$ is very
small, its contribution to $\left\langle A\right\rangle$ is also
very small. Then its corresponding state $u^\alpha$ can be
discarded. Using this method, some states can be discarded during
the growth of the system size and the size of the Hamilitonian of
the system to be calculated is therefore reduced.

There are two basic DMRG algorithms---the infinite system and the
finite system algorithms. For the infinite system case, the main
process is as follows. We first choose a small-size system that can
be exactly diagonalized, e.g. $L=4$ sites is taken in our
calculations, as the superblock. Then use its ground state to form
the reduced density matrix $\rho$ of the system block. The $m$
highest eigenvectors of $\rho$ are kept to renormalize the
Hamilitonian of the system block $S$ and the corresponding
operators. We add then two new sites and use these renormalized
Hamiltonian and operators of $S$ to form a new superblock. By
repeating these steps, the system size grows but the size of the
Hamiltonian of the superblock keeps on a suitable size. 

The finite-system case is based on the infinite system case by sweeping the
superblock to reach a higher accuracy. For further detailes on the process,
see Ref.~\cite{SRW99}.

In our DMRG calculations we have used the finite-system DMRG algorithm.
Three sweeps have been taken to increase the accuracy. The numerical
calculations were performed on finite chains, up to $100$ lattice
sites, using the open boundary condition. Two sites were added in each
step. For accuracy, the largest kept-state number reached $130$ and
the truncation error is less than $10^{-7}$. To avoid the influence
of the edge effect, 
the $r=0$ point was chosen in the middle of the chain.


\section*{References}

\begin{thebibliography}{99}

\bibitem{npbIII} J.M.P. Carmelo, arXiv:1211.5391; 
J.M.P. Carmelo and P.D. Sacramento, arXiv:1211.6073. 


\bibitem{Carmelo_2004} J.M.P. Carmelo, J. Rom\'an and K. Penc, Nucl. Phys. B {\bf 683},
387 (2004).

\bibitem{Harris_1967} A.~B. Harris and R.~V. Lange, Phys. Rev.
{\bf 157}, 295 (1967).

\bibitem{Stein_1997} J. Stein, J. Stat. Phys. {\bf 88}, 487 (1997).

\bibitem{bipartite} J.M.P. Carmelo, S. \"Ostlund and M. J. Sampaio, Ann. Phys. {\bf 325}, 1550 (2010).

\bibitem{macdonald} A.~H. MacDonald, S.~M. Girvin and D. Yoshioka,
Phys. Rev. B {\bf 37}, 9753 (1988).

\bibitem{Leigh} R.G. Leigh, P. Phillips and T.-P. Choy, Phys. Rev. Lett. {\bf 99}, 046404 (2007). 

\bibitem{Choy} T.-P. Choy, R.G. Leigh, P. Phillips and P.D. Powell, Phys. Rev. B {\bf 77}, 014512 (2008). 

\bibitem{Phillips} P. Phillips, T.-P. Choy and R.G. Leigh, Rep. Prog. Phys. {\bf 72}, 036501 (2009). 

\bibitem{Kaplan} T.A. Kaplan, P. Horsch and P. Fulde, Phys. Rev. Lett. {\bf 49}, 889 (1982). 

\bibitem{Montorsi} A. Montorsi and M. Roncaglia, arXiv:1207.3426.

\bibitem{Zou_1988}  Z. Zou and P.~W. Anderson, Phys. Rev. B {\bf 37},
627 (1988).

\bibitem{Ostlund_2006} S. \"{O}stlund and M. Granath, Phys. Rev. Lett.
{\bf 96}, 066404 (2006).

\bibitem{Affleck} I. Affleck, D. Gepner, H. J. Schulz, and T. Ziman, J. Phys. A {\bf 22}, 511 (1989).

\bibitem{Giamarchi} T. Giamarchi and H. J. Schulz, Phys. Rev. B {\bf 39}, 4620 (1989). 

\bibitem{Singh} R. R. P. Singh, M. E. Fisher, and R. Shankar, Phys. Rev. B {\bf 39}, 2562 (1989). 

\bibitem{Hallberg} K. A. Hallberg, P. Horsch, and G. Mart\'{\i}nez, Phys. Rev. B {\bf 52}, R719 (1995). 

\bibitem{Hikihara} T. Hikihara and A. Furusaki, Phys. Rev. B {\bf 58}, R583 (1998).

\bibitem{kotliar} G. Kotliar and A.E. Ruckenstein, Phys. Rev. Lett. {\bf 57}, 1362 (1986).

\bibitem{dorin} V. Dorin and P. Schlottmann, Phys. Rev. B {\bf 47}, 5095 (1993).

\bibitem{coleman} P. Coleman, Phys. Rev. B {\bf 29}, 3035 (1984).

\bibitem{ricardo} R.G. Dias and J.M.B. Lopes dos Santos, J. Physique I {\bf 2}, 1889
(1992).

\bibitem{stiffness} N.M.R. Peres, R.G. Dias, P.D. Sacramento and J.M.P. Carmelo,
Phys. Rev. B {\bf 61}, 5169 (2000).

\bibitem{u1paper} P. Ribeiro, P.D. Sacramento and M.A.N. Ara\'ujo, Ann. of Phys.
{\bf 326}, 1189 (2011).

\bibitem{Lee_2006} P.~A. Lee, N. Nagaosa  and X.-G. Wen,
Rev. Mod. Phys. {\bf 78}, 17 (2006).

\bibitem{shiba72}
H. Shiba, {\it Prog. Theor. Phys} {\bf 48 }, 2171 (1972);
C. N. Yang and S. C. Zhang, Mod. Phys. Lett. B 4, 759 (1990);
Shoucheng  Zhang, Phys. Rev. Lett. 65, 120 (1990);
V. J. Emery, Phys. Rev. B 14, 2989 (1976);
A.B. Eriksson, T. Einarsson and S. \"Ostlund,  Phys. Rev. B  {\bf 52} , 3662 (1995)

\bibitem{majoranas} F.A. Berezin and M.S. Marinov, JETP Lett. {\bf 21}, 320 (1975); Ann. Phys., N Y 
{\bf 104}, 336 (1977);
V.R. Vieira, Phys. Rev. B {\bf 23}, 6043 (1981).

\bibitem{Carmelo_2010} J.M.P. Carmelo, Nucl. Phys. B {\bf 824}, 452 (2010).

\bibitem{wootters} W.K. Wootters, Phys. Rev. Lett. {\bf 80}, 2245 (1998).

\bibitem{SRW99}
S. R White, et al., Density-Matrix Renormalization: A New Numerical
Method in Physics, Springer, Berlin, (1999).



\end{thebibliography}
\end{document}